\documentclass[]{aa}  

\usepackage{tabularx}
\usepackage{tablefootnote}
\usepackage{threeparttable}
\usepackage[figuresright]{rotating}
\usepackage{txfonts}
\usepackage{graphicx}
\usepackage{nicefrac,xfrac}
\usepackage{upgreek}
\graphicspath{{images}{figs}}
\usepackage[breaklinks=true]{hyperref}
\usepackage{mathrsfs}
\usepackage{tikz}
\usepackage{nicefrac}
\usepackage{orcidlink}

\begin{document} 

   \title{Properties of Giant Radio Galaxies larger than 3 Mpc}
   \authorrunning{H. Andernach et al.}
	\author{H. Andernach \inst{1,2\orcidlink{0000-0003-4873-1681}}
 \and
 M. Br\"uggen\inst{3\orcidlink{0000-0002-3369-7735}}}
 
\institute{Th\"uringer Landessternwarte, Sternwarte 5, D-07778 Tautenburg, Germany\\
    \email{heinz@ugto.mx}
\and 
Permanent address: Depto. de Astronom{\'{i}}a, DCNE, Univ. de Guanajuato, Callej\'on de Jalisco s/n, C.P. 36023 Guanajuato, Mexico
\and
Hamburg Observatory, University of Hamburg, Gojenbergsweg 112, 21029 Hamburg, Germany}

   \date{Received: 2024; accepted: 2024}

  \abstract
  {Giant radio galaxies (GRG) are radio galaxies with physical sizes of their radio emission
  larger than 0.7 Mpc. Recently, the sample of GRGs has become large enough to study the extreme end of the GRG size distribution.}
  {We examine the properties of GRGs with largest linear sizes larger than 3~Mpc in order to shed light on the nature and origin of GRGs.}
  {We select, corroborate, and revise if necessary, the largest GRGs
  from literature. We add to these the GRGs we identified in our own search
  in modern radio survey images, combined with optical surveys and
catalogues of spectroscopic and photometric redshifts in order to find their
projected linear radio size.  We study their radio power--size relation,
the asymmetry in the lobes, their association with clusters of galaxies, as well
as their bending angles.}
{We present an unprecedented sample of 143 GRGs larger than 3~Mpc, 
of which 69 were newly found by us. The sample includes six GRGs with projected 
linear sizes clearly exceeding 5~Mpc and reaching up to 6.6~Mpc.  We find that
GRGs larger than 3~Mpc are distributed in redshift and radio luminosity 
indistinguishable from those of smaller GRGs. The same applies
to the fraction of quasars among their hosts. 
At most a single one of the GRGs larger than 3 Mpc can be classified as a 
clear Fanaroff-Riley (FR) type~I source, and only 6 per cent deviate from a 
clear FR\,II radio morphology.
One quarter of our GRGs show very diffuse lobes typical for remnant radio
galaxies, and only 59 per cent show indications of hotspots in at
least one lobe, with 38 per cent featuring a hotspot in both lobes.
As in the case of
smaller radio galaxies, the shorter lobe is most often also the brighter one.
We find tentative evidence that the bending angle decreases with size
of the GRG, but no trend with redshift is detected. 
The fraction of GRGs > 3 Mpc associated with catalogued clusters of galaxies
is around 15 per cent and thus indistinguishable from that of smaller GRGs. 
The bending angles of GRGs > 3 Mpc in known clusters are larger than for 
those GRGs not associated with clusters.
}
  {}


   \maketitle
%

\section{Introduction}
\label{sec:intro}

Radio galaxies (RG) come in a wide range of sizes. The largest ones have come to be called giant radio galaxies (GRGs), which typically denotes RGs with largest (projected) linear sizes (LLS) in excess of 700~kpc \citep{Willis1974, Barthel1985, Kuzmicz2018,2023JApA...44...13D}. They are relatively rare with a number density of around 5 per volume of (100~Mpc)$^3$ \citep{2023A&A...672A.163O} or a sky density of around 3 per square degree \citep{Simonte2023} in the currently
deepest fields observed. GRGs are important to study the intergalactic medium and the evolution of radio jets. Their existence suggests that radio jets can maintain power over a few tens of Megayears and avoid disruption by interactions with the ambient medium, even when the Universe was around ten times denser than it is today.
GRGs may also play an important role in magnetising intergalactic space \citep{1986A&A...169...63K,1999MNRAS.309..100I,2023MNRAS.519.5723O}. 
In a recent paper, \cite{2024Natur.633..537O} note that a single void-penetrating lobe could fill its void with a magnetic field of strength $B \sim 10^{-16} - 10^{-15}$ G within a Hubble time.

Numerous radio surveys conducted in the past, such as the Faint Images of the Radio Sky at Twenty-cm \citep[FIRST,][]{FIRST1995}, the Westerbork Northern Sky Survey \citep[WENSS,][]{WENSS1997}, the National Radio Astronomy Observatory (NRAO) VLA Sky Survey \citep[NVSS,][]{1998AJ....115.1693C}, the Sydney University Molonglo Sky Survey \citep[SUMSS,][]{SUMSS2003}, and the recent surveys with the Australian Square Kilometre Array Pathfinder
(ASKAP), like the Evolutionary Map of the Universe \citep[EMU,][]{2021PASA...38...46N},
the Rapid ASKAP Continuum Survey \citep[RACS,][]{2020PASA...37...48M}, have uncovered around 1000 GRGs \citep{Ishwara-Chandra1999, Lara2001, 2001A&A...374..861S, Machalski2001, Saripalli2005, Kuzmicz2012, Kuzmicz2021, Kuzmicz2018, Dabhade2017, Dabhade2020a, Bruggen2021, 2021Galax...9...99A, Gurkan2022, Mahato2022, 2021MNRAS.501.3833D}. In addition, the LOw-Frequency ARray \citep[LOFAR,][]{LOFAR2013}, due to its high resolution and sensitivity to sources with very low surface brightness, marks the beginning of a new era in the study of very large and high-redshift RGs. The LOFAR Two-metre Sky Survey \citep[LoTSS][]{Shimwell2019, 2022A&A...659A...1S} has led to the identification of approximately 10,000 GRGs (\citealt{2020A&A...635A...5D, Simonte2022, 2023A&A...672A.163O}, \citealt{2024A&A...691A.185M}). 

\cite{Simonte2024} found differences in the environmental densities between GRGs and smaller RGs. 
They could show that GRGs are preferentially found in sparser environments compared to smaller RGs, and presented evidence that larger sources exhibit steeper integrated spectral indices, suggesting that GRGs are late-stage versions of RGs.
While the majority of GRGs are known to inhabit underdense environments, \cite{2020A&A...635A...5D} 
and \cite{2021Galax...9...99A} had found that between 10 and 20 per cent of GRGs are located in clusters, and recently \citet{2024A&A...687L...8S} found an even higher percentage of $\sim$24 per cent, 
as well as a smaller average linear size of GRGs in clusters.

Despite these recent advancements, the origin of the Mpc-sized GRGs remains unexplained. Existing models of RGs suggest that only a limited set of conditions involving jet power, environmental factors, host galaxy characteristics, and source age can account for the immense size of GRGs \citep{Turner2015, Hardcastle2018, Turner2023}. 

The large progress made in discovering GRGs has led us to take a look at the extreme end of the GRG population, focussing on GRGs with LLS $>$ 3 Mpc. This sample currently comprises 143 GRGs, including six that have LLS $>$ 5 Mpc,
discussed here in some detail for the first time. 
We note that there is no physical reason for adopting a threshold of
0.7~Mpc for the definition of GRGs, nor for our threshold of 3~Mpc.
With the latter value we have tried to select the largest possible 
sample of the physically largest GRGs for which it is still manageable 
to measure the parameters we studied. Our sample corresponds to about
1 per cent of all known GRGs larger than 0.7~Mpc, motivating us to 
search for features that would distinguish such extreme GRGs from 
the smaller ones.

The outline of this paper is as follows: In Sec.~\ref{sec:sample} we describe our sample. In Sec.~\ref{sec:results} we present our results, in particular looking at the power-size diagram, the asymmetry and the bending angles. In Sec.~\ref{sec:disc} we discuss the implications and conclude. Throughout this paper we use standard cosmological parameters, with
 $H_0=70$~km/s/Mpc, $\Omega_m = 0.3$ and $\Omega_{\Lambda}=0.7$, and we define the
 radio spectral index $\alpha$ according to $S_{\nu}\propto\nu^{\alpha}$. Position angles
 (PA) on the sky are quoted from North through East.

\section{The Sample of GRGs}
\label{sec:sample}

Over the past 12 years one of us (H.A).
\citep[see e.g.,][]{2012sngi.confP...1A, 2018arXiv180807178V}  
has compiled a growing list of extended RGs
with information on their exact host position, host name, apparent
magnitude, redshift, angular and linear size of the radio emission and a crude
radio-morphological classification. The list comprises all GRGs from
many dozens of papers from the published literature, which are then 
verified and corrected or discarded when necessary, but its major content
derives from an independent visual inspection of radio survey images at 
various frequencies and angular resolutions, combined with
optical or mid-infrared surveys to identify the host objects. 
For the present paper we also recently inspected large parts of the
Rapid ASKAP Continuum Survey \citep[RACS,][]{2020PASA...37...48M} at 
888\,MHz as well as all 651 MeerKAT pointings from the MeerKAT Galaxy Cluster Legacy Survey \citep[MGCLS,][]{2022A&A...657A..56K}, the 
MeerKAT Absorption Line Survey (MALS) DR2 \citep{2024A&A...690A.163W},
and the MeerKAT follow-up of GLEAM 4-Jy (G4Jy) sources by \cite{2023MNRAS.518.4290S}.
The recent completion of all three epochs of the Very Large Array Sky Survey \citep[VLASS,][]{2020PASP..132c5001L}, \url{http://cutouts.cirada.ca} as well as the recent release of the median stack of these at
\url{http://archive-new.nrao.edu/vlass/HiPS/MedianStack/Quicklook}
has also helped in finding or revising the hosts of previously known GRGs,
at least for those at DEC$>-40^{\circ}$.
For our method to measure
the LAS on radio survey image we refer to Sec.~2 of \citet{Simonte2022}.
A growing number of, both, spectroscopic and photometric redshift catalogues \citep[see][]{Simonte2024} is then used to
find the distance to the hosts and convert their ``largest angular radio size'' (LAS) to their (projected) linear size (LLS). 
Currently the compilation contains 28,700 extended
radio sources with a median LAS of 2.0$'$, median redshift of 0.47, and
a median LLS of 0.61\,Mpc, with over 12,000 GRGs larger than 700\,kpc. 
Note that our LLS values denote the projected lengths on the plane of the sky. 
Since most GRGs will not lie in that plane, \cite{2023MNRAS.518..240O} have used a statistical deprojection  algorithm to approximate the real physical
lengths. We refrain from doing this here for the uncertainties involved.

\begin{figure}
\centering
\includegraphics[width=1\columnwidth]{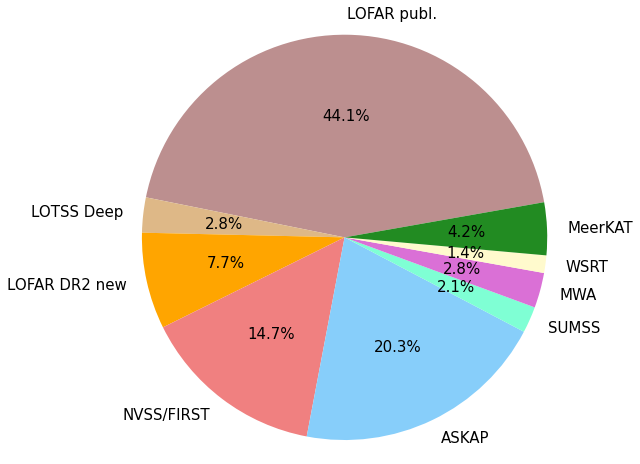}
  \caption{Provenance of the GRGs larger than 3~Mpc. Counterclockwise from top 
  we have:
  LOFAR publ.: found and published from LoTSS \citep{2020A&A...635A...5D,2023A&A...672A.163O,2023A&A...674A.208M}, LoTSS Deep \citep{Simonte2024,2024A&A...683A.179B}, and ``LoTSS DR2 new'': our own 
  discoveries from LoTSS DR2 images,
  the National Radio Astronomy Observatory (NRAO) VLA Sky Survey \citep[NVSS,][]{1998AJ....115.1693C} and
  Faint Images of the Radio Sky at Twenty-cm \citep[FIRST,][]{FIRST1995}, 
  the recent surveys with the Australian Square Kilometre Array Pathfinder 
  \citep[ASKAP,][]{2020PASA...37...48M,2021PASA...38...46N}, 
  the Sydney University Molonglo Sky Survey \citep[SUMSS,][]{SUMSS2003},  the Murchison Widefield Array surveys \citep[MWA,][]{2016MNRAS.458.1057O,2017MNRAS.464.1146H,2022PASA...39...35H,2021PASA...38...57L}, 
  the Westerbork Northern Sky Survey \citep[WENSS,][]{WENSS1997}obtained with
  the Westerbork Synthesis Radio Telescope (WSRT), and the MeerKAT MALS DR2 \citep{2024A&A...690A.163W}.}
     \label{fig:pie}
\end{figure}

In Fig.~\ref{fig:pie}, we show the provenance of the GRG discoveries in terms
of the radio surveys in which they were first seen. Taking together the 
published GRGs and our own yet unpublished findings, LOFAR has provided 
55 per cent of the GRGs in our sample, with the ASKAP 
surveys providing the second-largest fraction of 20 per cent of GRGs.

The median redshift of our sample is $z_{\rm med}=0.67$, where 33 per cent of the redshifts are spectroscopic (see Table~1). 16 per cent of the GRGs are either spectroscopically confirmed 
quasars or quasar candidates (defined here as hosts with either
stellar appearance and/or WISE colours typical for quasars). This fraction 
is very similar for samples of GRGs in the LLS range between 1 and 3 Mpc. 
For example, for the 180 GRGs listed in \cite{2021Galax...9...99A} with LLS from 1 to 3~Mpc, the median redshift is $z_{\rm med}=0.55$ and median power 
$\log P_{\rm 145 MHz}/{\rm (W/Hz)} =26.3$,
with 15 per cent of these being quasars. For the 128 GRGs LLS from 1 to 3~Mpc in the 
LoTSS deep fields found by \cite{Simonte2024}  these values are $z_{\rm med}=0.76$, $\log P_{\rm 145 MHz}/{\rm (W/Hz)} =25.7$ and 13 per cent, respectively. Since the latter is the
highest-sensitivity radio survey yet searched for GRGs, it is not surprising that
their median redshifts are higher, and their median radio power is lower.

\begin{figure}
\centering
\includegraphics[width=1\columnwidth]{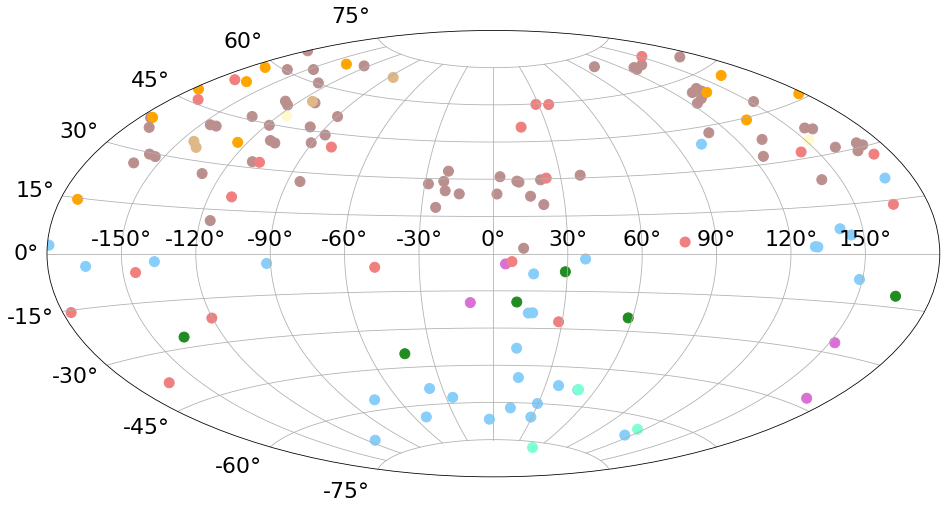}
  \caption{Sky distribution of our sample in equatorial coordinates, with the same color coding as in Fig.~\ref{fig:pie}. }
     \label{fig:sky}
\end{figure}

In Fig.~\ref{fig:sky}, we show the sky distribution of all 143 GRGs, 
where 65 per cent of the sources lie in the northern sky. Only 12 GRGs 
lie within 20.3$^{\circ}$ from the Galactic plane, and all but one of
the latter, J1821+2639, were newly discovered by us.

While GRGs lie preferentially in underdense regions \citep{Simonte2024}, no relation
between LLS and ambient galaxy has been found \citep{Komberg2009, Lan2021}. In fact,
a fraction of GRGs are also known to lie in overdense regions, such as in 
clusters of galaxies and even brightest cluster galaxies (BrClG's in what follows), 
and \citet{2020A&A...635A...5D}
and \citet{2021Galax...9...99A} have shown that between 10 per cent and 20 per cent of GRGs reside in clusters or groups. It is difficult to study this for our full GRG sample as it has a median $z\sim 0.67$ which
implies that only for less than half of these GRGs a cluster at their redshift could be seen on current deep optical survey images. Nevertheless, in Sect.~\ref{sec:clusters} we show that even with currently available data the cluster association fraction is the same for GRGs larger than 3~Mpc and those smaller than this.

\section{Results}
\label{sec:results}

\subsection{$P-D$ Diagram}
\label{sec:pd}

It is instructive to plot the radio power against the linear size in a so-called $P-D$ diagram \citep[$P-D$ diagram,][]{Baldwin1982}. It has been suggested that every radio source has a specific evolutionary track in this diagram which depends on source properties such as jet power or the environment \citep[see][]{Ishwara1999, Machalski2004, Hardcastle2018}. 

However, radio powers are not readily available for most GRGs of any size,
since this requires careful integration of their total flux densities from
the best available survey images. We did this for our sample using the
\texttt{draw} option of the Aladin software \citep{2000A&AS..143...33B},
correcting the resulting pixel sums by the ratio of beam-to-pixel area
of the respective survey images used, and comparing, where possible, 
with source catalogues available for these surveys. The integration areas were
drawn by hand, avoiding sources supposed to be unrelated to the GRG.
We note that a significant fraction of GRGs have lobes of very low
surface brightness for which total flux densities may have uncertainties
of more than 50 per cent. From the distribution of the derived radio
powers we are nevertheless confident that our results are reasonable,
but we refrain from quoting quantitative error values. Since most
GRGs were found in LoTSS at 145~MHz, the fluxes we derived on the best 
available survey images were then converted from the respective survey 
frequency to 145~MHz assuming a spectral index of $-$0.8.

\begin{figure}
\centering
\includegraphics[width=1\columnwidth]{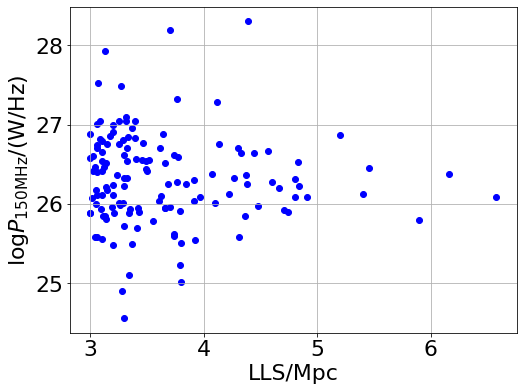}
  \caption{Power-size diagram for 140 of our 143 GRGs. The logarithm of the 
  radio power is given at a frequency of 145 MHz. Measurements at other frequencies 
  are extrapolated using a spectral index of $\alpha=-0.8$.}
     \label{fig:pd}
\end{figure}

We show the position of our GRGs sample in the $P-D$ diagram in Fig.~\ref{fig:pd}.
If GRGs are old radio galaxies, it is expected that they are  less powerful than smaller radio sources. Hence, there are no GRGs in the upper-right corner of Fig.~\ref{fig:pd}, where very powerful and large RGs should reside. For a giant source, the integrated signal-to-noise is proportional to the integrated flux divided by the angular size. Applying this to the $P-D$ diagram, the radio power, $P$, on the $y$-axis is proportional to the integrated surface brightness and the LLS, on the $x$-axis, is proportional to the angular size (all neglecting redshift effects). This means that the lines of constant signal-to-noise increase linearly with LLS which explains the dearth of GRGs in the bottom right corner.

In particular, we draw attention to the four high-luminosity outliers with $\log P/{\rm (W/Hz)} >$ 27.5 and the two low-luminosity outliers
at $\log P/{\rm (W/Hz)} < 25$
that lie outside the bulk of sources. Here we do not try to model their powers 
in the framework of evolutionary models but we suspect this may prove difficult
unless they have very high powers when the sources were smaller. These sources 
are interesting cases for source models and may lead to constrain additional 
mechanisms such as re-acceleration or other processes. In what follows we discuss the four overluminous GRGs in order of 
increasing LLS, followed by the one underluminous GRG.

J1102+1345 has a very regular FR\,II structure in RACS\-low \citep{2021PASA...38...58H}, RACSmid \citep{2023PASA...40...34D}, and VLASS \citep{2020PASP..132c5001L},
with a clear radio core detected. The $z_{\rm spec}$=0.671 was measured by
us based on a spectrum obtained with the 10-m Gran Telescopio de Canarias
(GTC). The Eastern hotspot has the bluish galaxy SDSS~J110304.71+134542.9
superposed which is very faint in WISE and for which LAMOST
(\url{http://dr6.lamost.org}) quotes $z_{\rm spec}$=0.059.  The VLASS image
shows the hotspot emission clearly displaced from this galaxy, thus we
have no evidence that this galaxy contributes to high radio luminosity
of this GRG.

J0101+5052 is located at $b=-12^{\circ}$ Galactic latitude (extinction of
$A_r=0.5$\,mag). VLASS images show two compact sources along
the major axis of the source. We prefer the optically fainter Southern of the
two (listed in Table~1) as it has a higher VLASS flux and more AGN-
or QSO-like WISE colors.  The optically brighter northern of the two
coincides with the Pan-STARRS object \citep{2020ApJS..251....7F} PSO~J015.4837+50.8932 
for which we found two photometric
redshifts \citep{2021MNRAS.500.1633B,2022MNRAS.515.4711B}
with mean of $z_{\rm phot}$=0.225 which would lead to an LLS of
1.32~Mpc, and to a moderate $\log P_{\rm 145 MHz}/{\rm (W/Hz)} =26.0$.

J1558$-$2138 is located at $b=+23.4^{\circ}$ Galactic latitude (extinction
of $A_r=0.7$\,mag) and was observed as G4Jy~1289 with MeerKAT by
\citet{2023MNRAS.518.4290S} who do not propose a host. The source is
very diffuse but both the MeerKAT and VLASS images show a point source
coincident with the host we propose in Table~1. No photometric redshift
is available for it in the literature, but we consider our estimate of 0.55 as
conservative in view of its faintness ($i=21.47$ in the DESI Legacy Survey
at \url{https://www.legacysurvey.org}). The source is the most complex
and the one with the highest (though difficult to measure) bending angle.

However, \citet{2025MNRAS.537..343Q} have this source as a remnant RG
identified with 2MASX~J15590200-2140032 (while the authors do not say this
explicitly, this is the only object that complies with the redshift search
method they describe in their paper). This is the brightest galaxy
near the geometric center of the source and is faintly detected in
RACS-low \citep{2020PASA...37...48M} and RACS-mid \citep{2023PASA...40...34D}
and possibly in the MeerKAT image by \cite{2023MNRAS.518.4290S}, but
is 10 times fainter in VLASS than the host proposed in Table 1.
We therefore consider 2MASX~J15590200-2140032 a less likely host, but
cannot exclude this possibility which would make the LLS of this
source a more modest 0.87~Mpc. Given its large bending angle, its removal
from the sample would not change our conclusions, and in fact make
the null hypothesis of no change in B.A. as function of LLS above
3~Mpc slightly more likely.

J0226+5927 is located at $b=-1.2^{\circ}$ Galactic latitude (extinction
of $A_r=2.3$\,mag) and was described by \citet{2021MNRAS.500.3111B}
within a soft $\gamma$-ray selected GRG sample where it appears as
SWIFT~J0225.8+5946. The host is 
also detected in the optical as PSO~J036.6064+59.4640 with $i_K$=20.63. 
We estimate its
redshift based on its QSO-like WISE colors and use Fig.~2 of
\citet{2018ApJS..235...10K} to infer $z\sim1.3$.

The two sources with the lowest radio power are: 

J0911+5440 with $\log P_{\rm 145 MHz}/{\rm (W/Hz)} =24.6$ has an
inner double-jet feature of 1.4$'$ length and detached, very low surface 
brightness lobes, similar to those of J0814+5224
\citep[aka ``Alcyoneus'', c.f.][]{2022A&A...660A...2O}. It is possible
that the total flux of these lobes, and thus its radio power, is higher 
than those determined from the LoTSS DR2 image.

J0955-0744 with $\log P_{\rm 145 MHz}/{\rm (W/Hz)} =24.9$ 
was found by L.\ Rudnick in a tile of the  EMU main survey (Hopkins et al.\ 2025, submitted).
It has a straight and regular FR\,II morphology and its low radio power is likely
due to the combination of the high sensitivity of EMU and the low redshift ($z_{phot}=0.185$)
of its host.

\subsection{GRGs larger than 5 Mpc}
\label{sec:5mpc}

Our sample contains six GRGs with LLS $>5$ Mpc, shown in Fig.~\ref{fig:5mpc}. Three of these are contained in the LOTSS paper by \cite{Hardcastle2023}, and the other three were found by us. 

These six GRGs are, in order of increasing LLS: J1108+2916 (LLS $> 5.2$ Mpc), J0901+4420  (LLS $>5.4$ Mpc), J1753-0332 (LLS=5.46\,Mpc,
J0103+2305 (LLS $ >5.9$ Mpc), J1529+6015 (LLS $> 6.16$ Mpc), and
J0838+5327 (LLS $> 6.58$ Mpc). All six are very straight with 
bending angles  of $\leq 3.8^\circ$. However, according to 
\citet{2022A&A...660A...2O} the
GRG ``Alcyoneus'' (J0814+5224) would be the next-largest one with LLS = 5.0 Mpc,
and a bending angle of 22$^{\circ}$ (see sect.~\ref{sec:ba}),
but according to our way of measuring the LAS we find LLS = 4.74~Mpc. Thus, it stands out with the largest bending angle of all GRGs with LLS > 3.7\,Mpc. Below we discuss 
the six GRGs larger than 5\,Mpc individually.

\subsubsection{J0103+2305}

This source was taken from sect. 10.3 of \citet{Hardcastle2023}
where it is tentatively identified with a very faint host
DESI~J015.8514+23.0696, r=22.33, SDSS~J010324.36+230410.7, WISEA
J010324.31+230410.8, which  coincides with an apparent strong radio
core in LoTSS DR2. This object has various inconsistent 
photometric redshifts of 0.166 \citep{Brescia2014}, 0.302 \citep{Ahumada2020},
1.07 \citep{Zhou2021} and 1.7 \citep{Duncan2022},
and given its QSO-like WISE colors and faint optical magnitude a redshift near 1 seems most
reasonable. However, with a LAS=24.1$'$ this would lead to LLS=11.6~Mpc,
or two times the size of the largest currently known GRG. Hence, we consider
this host as unlikely. Even though the radio core is slightly extended
along PA=111$^{\circ}$ in VLASS \citep{2021ApJS..255...30G}, i.e.\ close to the
orientation axis of the lobes, it is also listed as the blazar candidate
BROS~J0103.4+2304 by \citet{2020ApJ...901....3I} which also argues against
it being the host of a GRG. Our more detailed visual inspection of the
region revealed three optically brighter and less distant hosts,
which are (from east to west), SDSS~J010329.76+230454.7 (z$_{\rm phot}\sim0.3$),
SDSS~J010326.59+230457.9 (z$_{\rm phot}\sim0.2$), and SDSS J010324.53+230502.8
(z$_{\rm spec}=0.265$ from SDSS~DR16 \citep{Ahumada2020}).
We propose the latter, western-most of these, as the most likely host,
as it is the radio-brightest in LoTSS DR2 and the only detected one in
a stack of all three VLASS epochs (from \url{http://cutouts.cirada.ca/})
with flux densities of $\sim1.7$ and 0.4~mJy at 144\,MHz and 3\,GHz
giving a spectral index of $\alpha=-0.5$.
Taking $z=0.265$ as a lower limit for the (still to be confirmed) host
of this GRG, we find a minimum LLS of 5.9\,Mpc.

\subsubsection{J0838+5327}

We found this source in the LoTSS DR2 maps. It is a restarted
(or "double-double") GRG with an inner pair of lobes, clearly featuring a southern arm ending in a hotspot 3.2$'$ from the host. However, the
extent of the northern, inner arm is less clear. We propose the northern, inner hotspot
to be ILT~J083805.57+532841.7 located $\sim$1.6$'$ from the host because further
north there are three compact radio sources superposed near the major
radio source axis, namely SDSS J083811.03+533032.7 ($z_{\rm spec}=0.5386$
from SDSS~DR16) $\sim3.6'$ from the host, the late-type face-on galaxy
2MASX~J08381599+5332160 ($z_{\rm spec}=0.06247$ from SDSS~DR16) $\sim5.5'$ from
the host, and $\sim7.9'$ from the host there is ILT~J083822.18+533427.7,
the optical counterpart of which is hidden in the glare of a bright star
$\sim9''$ WSW of the radio source. The length ratios (brighter-to-fainter)
of the inner and outer doubles would then be 2.0 and 0.65, respectively,
though the outer lobes are very similar in total flux.

\subsubsection{J0901+4420}

This source was found by us in the LoTSS DR2 image. Its host has a
prominent 9-mJy radio core in VLASS, but is very faint ($r=25.3$) in
the DESI DR10 image.  The star Gaia 1009292060152667904 lies 3.2$''$
SE of the host, which confuses the WISE images. The only available
$z_{\rm phot}=0.87$ from \citet{Zhou2021} could even be
underestimated in view of the optical faintness of the object. With
$z=0.87$ its angular size of 11.7$'$ would yield a lower limit to its
LLS of 5.4\,Mpc.  The radio core is listed as slightly extended
along PA$\sim$130$^{\circ}$ in VLASS \citep{2021ApJS..255...30G}
and along PA$\sim$106$^{\circ}$ LoTSS DR2 catalogues \citep{2022A&A...659A...1S}, 
but is listed in FIRST \citep{2015ApJ...801...26H} with a small extent along
PA$\sim12^{\circ}$, i.e.\ closer to the radio axis at PA$\sim$130$^{\circ}$
of the outer lobes. The
core flux densities at 0.144, 1.4, and 3\,GHz are 22, 14, and 25 mJy,
respectively, which may indicate evidence for variability.  Half way
to southern lobe, the object DESI~J135.3744+44.2967, $r=23.65$, $z_{\rm phot}\sim1.0$
from \citet{Zhou2021,2021MNRAS.500.1633B,Duncan2022}
appears as a weak radio point source and we consider it an unlikely host
for the entire source given its asymmetric location. Along the northern lobe
the low surface brightness galaxy SDSS~J090124.11+442214.2 appears as
a superposed radio source.

\subsubsection{J1108+2916}

This GRG is listed by \citet{Hardcastle2023} with LLS=5.7\,Mpc. It
has rather symmetric lobes, gradually increasing in radio surface brightness from
the host to the outer hotspots. The only unusual feature is a smooth, but
significant bending towards the south about 1.5$'$ before reaching the SW hotspot.
We measure a LAS of 11.6$'$ and obtain an average $z_{\rm phot}=1.05$ from
\citet{Zhou2021,Duncan2022} and \citet{2022RAA....22f5001Z},
while SDSS~DR12 \citep{2015ApJS..219...12A} and \citet{Brescia2014} clearly
underestimate the redshift with $\sim0.385$. Two more intermediate redshifts
of 0.77 and 0.98 appear in \citet{2021MNRAS.500.1633B} and \citet{2022MNRAS.515.4711B}.
We adopt a conservative lower limit of $z_{\rm phot}>0.8$ to estimate a lower limit
for the LLS of 5.2\,Mpc.

J1108+2916 is accompanied in the sky by two other RGs
with almost parallel orientation, though clearly in the foreground of
the large GRG, namely SDSS J110829.69+291736.8, $z_{\rm spec}=0.2147$
and LLS=0.28\,Mpc located $\sim1.4'$ NW of the host, and
2MASX~J11082297+2914281, $z_{\rm spec}=0.2123$, and LLS=0.45\,Mpc, located
$\sim3.3'$ SW of the host.

\subsubsection{J1529+6015}

\begin{figure*}[!ht]
\centering

\includegraphics[height=14cm]{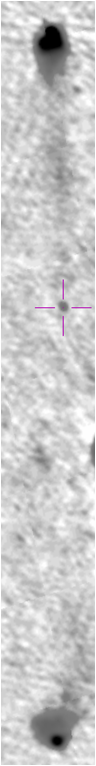}\hspace{0.5cm}
\includegraphics[height=14cm]{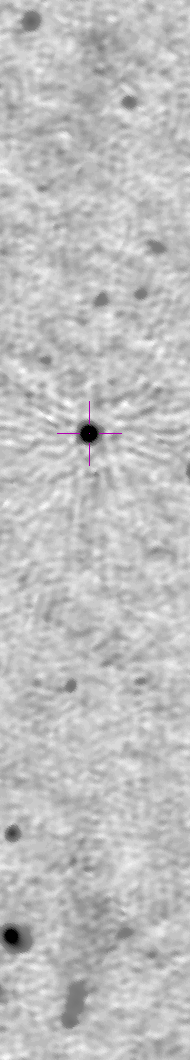}\hspace{0.5cm}
\includegraphics[height=14cm]{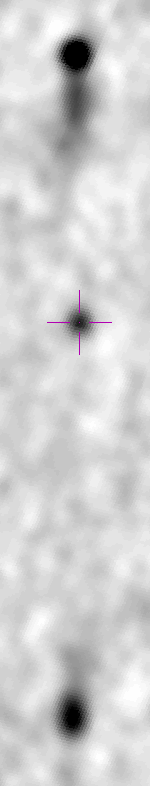}\hspace{0.5cm}
\includegraphics[height=14cm]{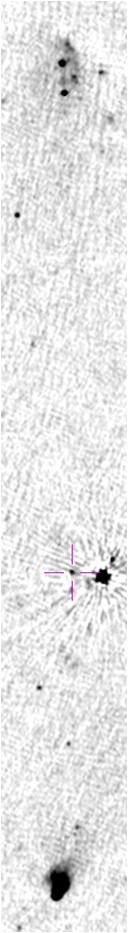}\hspace{0.5cm}
\includegraphics[height=14cm]{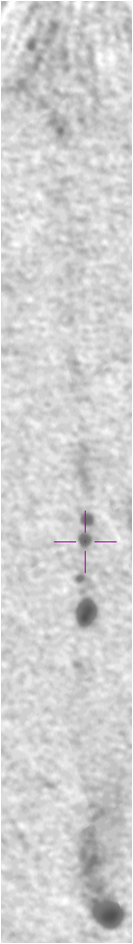}\hspace{0.5cm}
\includegraphics[height=14cm]{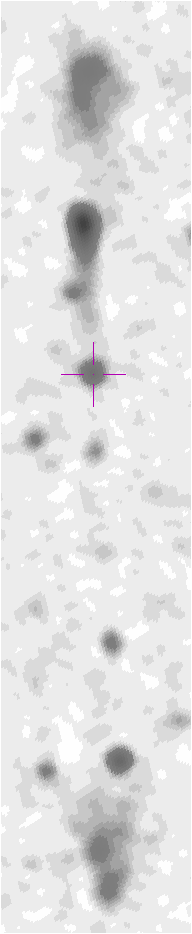}

\caption{The six known GRGs larger than 5 Mpc. Images from LoTSS DR2 are
placed in increasing order of their linear size from left to right, with
their host galaxy marked with a magenta cross. To save space, all
radio galaxies have been rotated to have their major radio axis vertical.
The GRGs are (with their LAS in arcmin, LLS in Mpc, and real PA on the sky,
from N through E, listed in parentheses): J1108+2916 (11.6$'$, >5.2, 31$^{\circ}$),
J0901+4420 (11.7$'$, >5.4, 170$^{\circ}$), J1753$-$0332 (13.6$'$, 5.46, 152$^{\circ}$), J0103+2305 (24.1$'$, >5.9, 101$^{\circ}$),
J1529+6015 (13.2$'$, 6.16, 26$^{\circ}$), and J0838+5327 (17.2$'$, 6.6, 22$^{\circ}$).
The lower limits for the LLS values are due to assuming the lowest reasonable
host redshift.}
     \label{fig:5mpc}
\end{figure*}

This GRG was presented in sect.\ 10.3 of \citet{Hardcastle2023} as having
an LLS of 6.9\,Mpc, albeit with some doubt about the correct host and
host redshift. The exact LAS of the source is somewhat uncertain due to
artefacts caused by a strong source hosted by SDSS J153004.29+602422.8 some
2.5$'$ north of the northern lobe of the GRG. We estimate LAS=17.2$'$ from the LoTSS~DR2
image. The most likely host is the galaxy SDSS~J152932.16+601534.4 (marked
with the magenta cross in our Fig.~\ref{fig:5mpc}), for which \cite{2024Natur.633..537O} 
(who called called this source ``Porphyrion'') recently
measured a $z_{\rm spec}=0.896$ resulting in LLS=6.16\,Mpc. However, at an angular distance of 19$''$ NNE of
this galaxy there is the radio-loud QSO SDSS~J152933.03+601552.5 with
$z_{\rm spec}=0.7984$ from SDSS, and confirmed by \cite{2024Natur.633..537O} with
$z_{\rm spec}=0.799$, which we assume to be superposed on the northern arm of this GRG.
If the QSO were the actual host of the GRG, its LLS would be 5.94\,Mpc.
It may be a restarted GRG, but in this case the inner double would be rather
asymmetric with the NE inner lobe ending in a very elongated radio ridge near
declination +60$^{\circ}$\,16.5$'$ and a diffuse double hotspot near declination +60$^{\circ}$\,14.5$'$.
Similar to the southern lobe of J1108+2916 the southern lobe of J1529+6015 bends westward about 1$'$ before reaching the southern hotspot.

\subsubsection{J1753$-$0332}

This GRG is clearly seen in the RACS-low image at 888\,MHz \citep{2020PASA...37...48M},
and a radio core in both RACS and VLASS confirms the host galaxy. Given its low 
Galactic latitude of +11.2$^{\circ}$ and $r$-band extinction of $A_r=1.7$\,mag, only a single
photometric redshift of $z_{\rm phot}=0.62\pm0.06$ from \cite{2021MNRAS.500.1633B} was found.
We have no reason to believe that this is an overestimate, and, moreover, the
presence of numerous faint, red galaxies within $\sim30$ arcsec of the host in the
Pan-STARRs DR1 color image \citep{2020ApJS..251....7F}, albeit without any
$z_{phot}$ value in \cite{2021MNRAS.500.1633B}, suggests the host to be
a member of a galaxy cluster lying in an area not covered by cluster catalogues.

\subsection{Radio Morphology and Asymmetry}
\label{sec:asym}

From a visual inspection of the shape of the radio emission in the best available 
survey images, we find that 25 per cent of our 143 GRGs show very diffuse lobes
typical for remnant radio galaxies, and only 59 per cent show indications of hotspots
in at least one lobe, and only 38 per cent show a hotspot in both lobes.
We classified the radio morphology of each lobe of our 143 GRGs according 
to their FR type \citep{1974MNRAS.167P..31F} and found no single GRG
that would conform to an FR\,I type on both sides, except possibly 
J1821+2639 from \cite{2023A&A...672A.163O} which unfortunately lies outside
the public LoTSS DR2 area for which images are unavailable to us.
Only nine GRGs showed lobes that deviate from the FR\,II type 
and we describe these sources in what follows.


J0235+3104 is a remnant-type double-lobed source with no indication
of radio jets or hotspots.

J0331$-$7713 appears with a FR\,II type N lobe and a very low surface
brightness S lobe without indication of a hotspot.

J0740$-$6647 was recognized in RACS\-low and shows symmetric, rectangular-shaped
lobes with sharp edges on both sides along their major axes, plus faint
radio emission beyond their radio brightness peak, the location of 
which formally still conforms to an FR\,II type on both sides.

J0831+6711 has a hybrid morphology with its NE arm of FR\,I and its
SW arm of FR\,II type, albeit without evidence for a hotspot.

J1129+2733 was published by \citet{2023A&A...672A.163O} but is not
in the footprint of the public LoTSS DR2 release. Its NVSS image
led \citet{2012sngi.confP...1A} (bottom right figure) to propose it as
a candidate GRG. The RACS-low survey image \citep{2021PASA...38...58H} barely shows
the outer lobes, but suggests an inner double of 2.5$'$. However, 
VLASS shows it to be just the inner part of the jets, without indication
of inner hotspots, thus we do not consider it a restarted RG.

1530+5243 in LoTSS DR2 has a diffuse N lobe without hotspot and formally 
of FR\,I type, while the S lobe practically only consists of a hotspot with
low surface brightness emission only marginally detected even in the
low-resolution LoTSS image.

J1558$-$2138 appears as a remnant-type double-lobed source already
in NVSS, and in RACSmid \citep{2023PASA...40...34D} and MeerKAT
\citep{2023MNRAS.518.4290S} shows a complex and bent radio structure,
reminiscent of a wide-angle tailed (WAT) source without indications of jets 
or hotspots.

J1651+3209 is a rather straight FR\,II RG with a wide inner double
of 4.9$'$ extent (or 66 per cent of its total size) and ending in hotspots,
while the outer lobes are of hybrid morphology with the NE and SW
arms of type FR\,II and I, respectively.

Finally, 1902$-$5605, recognized in RACS\-low and measured by us in the 
EMU main
survey, shows a faint jet due NW ending in a diffuse lobe of constant surface
brightness without hotspot.  The SE lobe starts with a radio knot
$\sim1.5'$ from the host and ends in a diffuse emission region, formally
of FR\,II type, but without hotspot.

Next, we wish to discuss the asymmetry of the lobes of GRGs in our sample.
Asymmetry in RGs can be produced by three effects: (i) the jet production at the AGN could already be asymmetric, (ii) orientation can cause asymmetry owing to the differing light travel times to the observer for each arm and (iii) the external medium could be anisotropic, presenting different resistance to the jets.  From the fact that most sources show the stronger lobe on the shorter arm, \cite{Lara2004} concluded that the ambient medium is the dominant cause for asymmetry in their sample of RGs. 

\begin{figure}
\centering
\includegraphics[width=1\columnwidth]{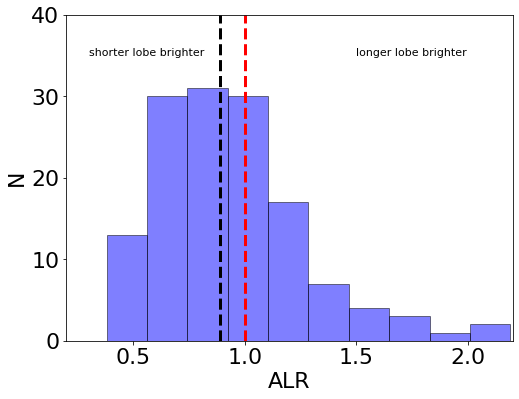}
\caption{Ratio of stronger-to-fainter lobe length for 140 of our 143 GRGs. An
ALR $>1$ implies that the longer lobe is brighter and ALR $<1$ that the shorter lobe
is brighter. The vertical red line at ALR=1 is drawn to guide the eye. 
The black line denotes the median at 0.89.}
     \label{fig:alr}
\end{figure}

The arm-length ratio, sometimes denoted as $r$, of a radio source is typically defined as the ratio between the length of its shorter to its longer arm. A value of $r$ near unity indicates a more symmetric jet ejection and propagation environment. \cite{Lara2004} found a clear difference between the $r$ distribution of FR~I and FR~II RGs. For FR~I types, the distribution gradually increases towards symmetry ($r$=1), with 50 per cent of the sources presenting $r>0.8$, while for FR~II types the distribution is more concentrated around a mean value of 0.78. They also look at the asymmetry of GRGs. Comparing the arm-length ratio of the GRGs of FR~II type with normal FR~II, they do not find that GRGs are more asymmetric than smaller RGs, while \cite{2000A&AS..146..293S}
noted that GRG asymmetries are slightly larger than those for smaller-sized RGs from 
the 3CR catalogue.
Here, we define the ALR as the (length of brighter lobe) / (length of fainter lobe). 
Thus, an ALR $>1$ implies that the longer lobe is brighter and ALR $<1$ that the 
shorter lobe is brighter. With this definition the ALR contains 
more information than the ratio $r$ described above.
We plot the ALR for 140 of our 143 GRGs in Fig.~\ref{fig:alr} and find that the 
ALR peaks at values below 1, around 0.85, which is similar to the distribution of a 
much larger
sample of smaller FR\,II sources drawn from the FIRST survey 
\citep[see Fig.~4 in][]{2019arXiv190809988D}.

When the lobe interacts with a denser environment, the jet head exerts more work, leading to a more efficient conversion of jet energy into radio emission \citep{Eilek1989,Blundell1999}. 

That side of the jet that propagates through a denser ambient medium will
tend to be shorter and may appear brighter than the other side. Hence, it
has been stated that jets with ALR < 1 provide evidence of an anisotropic
environment. \citet{Arshakian2000} found that asymmetry is more 
common in luminous sources.  However, we do not see any relation between 
radio power and ALR in our sample of GRGs. In a related study, \cite{Dabhade2024} found that cocoons in which the inner lobes of giant double-double RGs expand often contain dense material, causing an asymmetry of the inner lobes.

\subsection{Association with galaxy clusters}
\label{sec:clusters}

In this section, we will investigate the possible association of the GRGs in our sample 
with known galaxy clusters drawn from the deepest large-scale optical surveys like the
Sloan Digital Sky Survey 
\citep[SDSS,][]{2010ApJS..191..254H,2011ApJ...736...21S,2015ApJ...807..178W}, 
the Dark Energy Survey \citep[DES~DR2,][]{2022RAA....22f5001Z},
the Dark Energy Survey Instrument Legacy Survey \citep[DESI,][]{2021ApJ...909..143Y, 2021ApJS..253...56Z} based on DR8, 
\citep[based on DESI~DR9]{2024MNRAS.531.2285Y, 2022RAA....22f5001Z} and
\citep[based on DESI~DR10]{2024ApJS..272...39W}, or the very deep 
Hyper Suprime-Cam Subaru Strategic Program \citep[HSC-SSP,]{2018PASJ...70S..20O,2022RAA....22f5001Z}.
Of our 143 GRGs 82 fall within the SDSS survey footprint, 20 
fall within the DES~DR2 region \citep{2022RAA....22f5001Z}, 
126 fall within DESI~DR10, and only 4 fall within the HSC-SSP wide
survey.
Most of these catalogues list the position of the brightest cluster galaxy (BrClG)
as the cluster centre, and some also offer a list of individual cluster member galaxies
which justifies a cross-match of our GRG host positions within a 2$''$ search radius.
There is another cluster catalogue by \citet{2021ApJ...909..143Y} based on DESI~DR8
which lists the number of members of each galaxy system, including those with
a single ``member'' galaxy, i.e.\ an isolated galaxy. 
A cross-match within 1\,Mpc projected distance on the sky plane
from any of the 133,036 clusters in MaDCoWS2~DR2 \citep{2024ApJ...976..186T}
did not yield any cluster association for the 47 GRGs of our sample
covered by this catalogue.
For several other large
cluster catalogues (including some based on X-ray or Sunyaev-Zel'dovich effect 
detections) that do not list the BrClG position as the cluster centre, we used
a wider match radius, but no further cluster associations were found.
In the last column of Table~1 we indicate whether we found a GRG host as 
BrClG (b), a member galaxy (m), or as isolated (i) in any of these catalogues. 
Note that 16 of the 18 cluster-associated hosts are listed as BrClG and 
only two additional ones as cluster members. We note that one of our 
GRG hosts is listed as ``bi", i.e.\ both a BrClG and a non-brightest cluster 
member in
different catalogues, two others are listed as ``mi", i.e.\ both a cluster
member and an isolated galaxy, and three further GRGs are even listed as ``bi",
i.e.\ both a BrClG and an isolated galaxy. 
We interpret this as being a consequence of the different clustering algorithms 
used by different authors, and that cluster association fractions of GRG 
hosts have to be taken with caution.

In order to estimate the fraction of GRGs associated with clusters we must take
into account the typical redshift limits for the respective cluster catalogues.
The most distant GRGs with BrClG or cluster member matches were found 
near $z\sim$0.85, and the most distant isolated GRG host galaxy from \citet{2021ApJ...909..143Y} was found at $z\sim$0.9. 
We therefore consider the latter redshift as the upper limit out to which
we can hope to find any cluster association of our GRG hosts.  There are 
25 of our 143 GRGs with $z \geq 0.9$, of which three lie at 
Galactic latitude $|b|\le20^{\circ}$, which leaves us with 
124-22=102 of the 143 GRGs that may have been detected as cluster member in 
DESI~DR10 by  \citet{2024ApJS..272...39W}. We may thus conclude that at least 
16 of 102 (or 16 per cent) of our GRGs are cluster members, and all except two 
are listed as the BrClG, the latter including four GRGs with LLS$>$4\,Mpc. 
While this is a similar percentage as for the smaller GRGs \citep{2020A&A...635A...5D,2021Galax...9...99A}, it is in contrast with
\citet{2024A&A...687L...8S} who suggest that beyond a linear size of $\sim$3\,Mpc
GRGs are not found to be associated with clusters. The latter may be due to
the smaller sample of GRGs used by these authors. Apart from these cluster-associated
GRGs, 28 of the 102 GRGs with $z<$0.9 appear as isolated in the cluster
catalogue of \citet{2021ApJ...909..143Y}, leaving more than half of our GRGs
without clear evidence of lying in either over- or underdense regions. Deeper optical surveys allow one to detect clusters of galaxies out to redshifts beyond $\sim$1.

Thus we could show that at least 16 per cent of the GRGs are located in 
currently catalogued clusters of galaxies. Hence
their sizes cannot be explained by an underdense environment, but possibly by a higher jet power, which in turn may be caused by a more efficient feeding of the central black hole of the host galaxies.

\subsection{Bending Angles}
\label{sec:ba}

In this section, we study the bending angles (B.A.) which we measured on the best
available survey images, using the \texttt{dist} option in the Aladin software,
by connecting the host galaxy positions to the brightest pixels in the lobes of the GRGs. Then the B.A. is the difference angle from parallel lobe orientation.

\begin{figure}
\centering
\includegraphics[width=1\columnwidth]{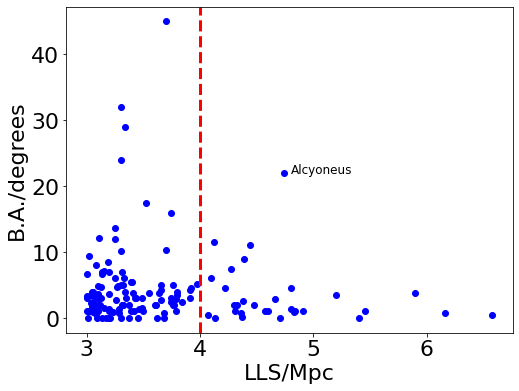}
  \caption{The bending angle, B.A., measured in degrees, for 140 of our 143 GRGs,
  plotted against the largest linear size, measured in Mpc. The location of GRG 
  J0814+5224 (Alcyoneus)
  in this plot is labelled \citep{2022A&A...660A...2O}. The vertical line separates the sample into GRGs with LLS$<$ 4~Mpc and $>$ 4~Mpc. Their median LLS is 3.4~Mpc.}
     \label{fig:ba}
\end{figure}

As shown in Fig.~\ref{fig:ba}, the B.A. appears to decrease with LLS. The median B.A. of the 109 GRGs with LLS between 3 and 4 Mpc is 3.0$^{\circ}$, while the median B.A. of the 31 GRGs with LLS larger than 4 Mpc is only 1.4$^{\circ}$. \\
Splitting the sample into equal sizes of 70 GRGs at LLS=3.39\,Mpc, we find
a median B.A. of 3.0$^{\circ}$ for the smaller GRGs and a median B.A. of 
2.8$^{\circ}$ for the larger ones. 
The student-t test yields a $p$-value of 0.186 and the KS-test a $p$-value of 0.88, which suggests that they belong to the same distribution.

None of several other statistical tests we performed did
suggest a significant difference in the B.A. distribution of GRGs with LLS between 3 
and 4\,Mpc and those larger than 4\,Mpc (e.g. $p=0.62$ for a KS-test). The apparent lack of B.A.s larger than $\sim10^{\circ}$ for GRGs >4\,Mpc 
may well be due to the small number of such large GRGs.
We did not find a significant trend for the B.A. in our sample to change with redshift.

If we compare the 17 GRG hosts listed as BrClG's (including the two listed as non-brightest cluster member) with the 27 isolated hosts, the B.A. distribution 
shows an interesting difference. The B.A. tends
to be larger for GRGs in clusters where higher ambient gas densities are expected to have a larger effect on radio lobes. The median B.A. in clusters is 3.5$^{\circ}$  while for the 22 isolated GRGs the median B.A. is 2.0$^{\circ}$. 
The student's t-test yields a low $p=0.033$ but the KS-test with $p=0.25$ suggests that 
this difference is statistically insignificant. Since the KS-test does not assume normality of the underlying distribution, it goes beyond testing for mean differences and may be more applicable here.
Moreover, we do not find any statistically significant differences for the value of ALR between cluster-associated and isolated galaxies. The median ALR for the
17 cluster-associated GRGs is 0.84, compared to 0.945 for the 27 isolated GRGs. 
The KS-test yields a $p$-value of 0.27, which suggests that they are possibly drawn from the same distribution.

\section{Discussion and Conclusions}
\label{sec:disc}

Our large collection of extremely large GRGs shows remarkable features. Most importantly, the energetics and the straight morphology raise a number of questions. Jets with lengths of around 5 Mpc can inject an energy into the surrounding medium that is comparable to the energy released during the merger of a galaxy cluster mergers \citep{2024Natur.633..537O}, and thus belong to the most energetic events since the big bang. The radio powers of our GRGs lie above the top line in figure 8 of \cite{2025arXiv250409303H} indicating total powers in excess of $10^{40}$ W. If this power is sustained for 100 Myr, this would yield a total energy of $> 10^{62}$ erg.

Thus, GRGs pose the following conundrum: In order to power them, the supermassive black hole needs to grow by a large fraction (e.g. an estimated $\sim 10^9 M_\odot$ for the case of J1529+6015 aka Porphyrion). Accretion of such masses is usually associated with the accretion of substructure that causes velocity dispersion in the collisional and non-collisional matter. Yet, the lobes of GRGs seem surprisingly straight. 

The straightness of the lobes suggests low peculiar velocities (${\rm v}_p<10^2$ km/s), which may be interpreted as the hosts being at the bottom of a local gravitational potential well, perhaps surrounded by a void. This explains the large LLS and the persistent collimation of the lobes as jets grow more resistant against Kelvin–Helmholtz instabilities as the ambient density decreases \citep{galaxies7030070}. Around 16 per cent of the GRGs lie in clusters, and preferentially hosted by
the brightest cluster galaxies, which often reside at the bottom of the gravitational wells where the intracluster medium is densest. 
However, clusters are dynamic environments. Also it is unclear how AGN feedback can operate efficiently, if the AGN at the centre of the cluster is a GRG that no longer participates in heating the cluster core. So far no simulations exist of jets that are this long, and we hope that our sample motivates numerical work in this area.\\

\begin{figure}
\centering
\includegraphics[width=1\columnwidth]{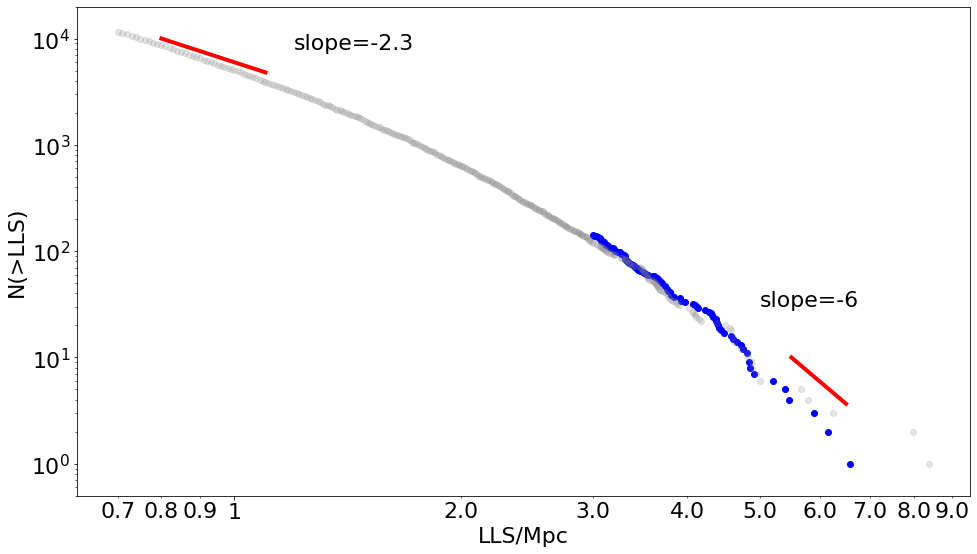}
  \caption{The number of known GRGs larger than a given LLS as function of LLS. Blue
  dots are for the 143 GRGs $>3$ Mpc from this publication, and grey dots are from \cite{2024A&A...691A.185M}. The lines of fixed slope in this log-log plot are shown for reference.}
     \label{fig:NLLS}
\end{figure}

Assuming that the intrinsic LLS follow a Pareto distribution, \cite{2023A&A...672A.163O} modelled projection and selection effects. 
They determined that the LLS were well described by a Pareto distribution with a tail index of $-3.5 \pm 0.5$. This implies that, given its projected length, a GRG’s intrinsic length is expected to be, on average, just 15 per cent larger. We fitted a Pareto distribution to our LLS distribution and found a higher tail index of -4.4, albeit not a very good fit. 
We show in Fig.~\ref{fig:NLLS} the number of GRGs larger than a given LLS from the present publication ($>3$ Mpc) in blue, and all smaller GRGs (in grey) 
from the compilation by \cite{2024A&A...691A.185M}. The discontinuity between blue and grey dots at 3~Mpc reflects our finding of more 
such GRGs, and the blue dots falling short of the grey ones for LLS$>$5.5~Mpc is 
due to our discarding of various GRGs from \cite{2024A&A...691A.185M}.
It is evident that the number of GRGs with LLS$>3$ Mpc starts to decrease 
sharply with slopes of $-$6 or steeper. 
This implies that these very long GRGs are very rare objects.\\

No fewer than about 19 (or 13 per cent) of the GRGs with LLS>3 Mpc lie at redshift $z\ge1$, although only 3 of these are spectroscopic (but for the others there are usually consistent photometric redshifts from more than one reference),
which raises the question how these could grow to these sizes with the conditions in the early Universe (matter density, CMB photon density) or be detected given cosmological surface brightness dimming.  
E.g., the QSO  SDSS~J004131.75+291217.6 hosts a 6.4$'$ wide FR\,II with diffuse lobes and has a spectroscopic redshift of $z=1.25$, yielding a factor of $(1+z)^4\approx25$ for the expected surface brightness dimming.
Most record-length outflows, such as in 3C\,236 (LLS = 4.6\,Mpc), J1420$-$0545 (LLS=4.84 Mpc) and J0814+5224 (Alcyoneus, LLS=5.0\,Mpc), are fuelled by radiatively inefficient AGN.\\

Recently, \cite{2025A&A...696L...8N} have inferred the magnetic field in the jets of Porphyrion GRG. Using the absence of gamma-ray emission, they determine a lower limit on the co-moving magnetic field strength of approximately 10 nG. This aligns with recent estimates of magnetic fields within the filaments of the cosmic web. Moreover, they explore the hypothesis that Porphyrion might be caused by a very-high-energy gamma-ray beam from an AGN. In this scenario, jets can maintain a straight appearance over vast distances without spreading into voids.\\

There have been discussions on the role of GRGs in magnetising the ambient medium \citep{2024Natur.633..537O}. In particular, one may wonder what fraction of the non-thermal components of the jets can diffuse out into the voids thus leading to wide-scale magnetisation. However, the persistent collimation of the jets may indicate that the diffusion of magnetic fields into the ambient medium may be limited.

The main conclusions of this paper are:

\begin{itemize}

\item The median redshift, median radio power, and quasar fraction in our sample 
of 143 GRG larger than 3~Mpc in projection are not different from that for the general sample of GRGs. \vspace{0.2cm}

\item None of our GRGs larger than 3~Mpc, with one possibly exception,
can be classified as a clear FR\,I-type source, and only 6 per cent deviate from a FR\,II radio morphology of both of their lobes. 
However, only 59 per cent show indications of hotspots
in at least one lobe, and 38 per cent show a hotspot in both lobes.  
About 25 per cent of our GRGs show very diffuse lobes typical of remnant radio galaxies. \vspace{0.2cm}

\item We find only marginal evidence that the B.A. decreases with LLS, but there is no discernible dependence on redshift.  \vspace{0.2cm}

\item As in the case of smaller radio galaxies, the shorter lobe is most often also the brighter one.  \vspace{0.2cm}

\item The most powerful GRGs in our sample appear difficult to reconcile with radio galaxy evolution models.  \vspace{0.2cm}

\item The fraction of GRGs > 3 Mpc associated with catalogued clusters of galaxies
is at least 16 per cent and thus indistinguishable from that of smaller GRGs. 
We find no evidence that the percentage of GRGs associated with clusters decreases with increasing projected radio size. \vspace{0.2cm}

\item The bending angles of GRGs > 3 Mpc in known clusters are larger than for 
 GRGs that are not associated with clusters.

\end{itemize}

\begin{acknowledgements}

HA is grateful to Hamburger Sternwarte, where part of this research 
was finalized, for their hospitality and financial support.
MB acknowledges funding by the Deutsche Forschungsgemeinschaft (DFG) under Germany's Excellence Strategy -- EXC 2121 ``Quantum Universe" --  390833306 and the DFG Research Group "Relativistic Jets". We are grateful to E.F. Jim\'enez Andrade,
R.F. Maldonado S\'anchez, A.\,Rodr{\'{i}}guez Y\'a\~nez, L.\,Rudnick, I.d.C. Santiago Bautista for having found some of the GRGs, and to A.\,Offringa and C.\,Lynch for providing images in which some GRGs were found. GTC spectra for 5 of the GRGs were reduced by I.\ Plauchu-Frayn
and R.\ Coziol.

This research has made use of the "Aladin sky atlas" developed at CDS, Strasbourg Observatory, France. We also acknowledge the use of the CIRADA cutout service at URL
\url{cutouts.cirada.ca}, operated by the Canadian Initiative for Radio Astronomy
Data Analysis (CIRADA). CIRADA is funded by a grant from the Canada
Foundation for Innovation 2017 Innovation Fund (Project 35999), as well
as by the Provinces of Ontario, British Columbia, Alberta, Manitoba and
Quebec, in collaboration with the National Research Council of Canada,
the US National Radio Astronomy Observatory and Australia’s Commonwealth
Scientific and Industrial Research Organisation.
This paper is based (in part) on data obtained with the International LOFAR Telescope (ILT) under project code LT10\_012. LOFAR (van Haarlem et al. 2013) is the Low Frequency Array designed and constructed by ASTRON. It has observing, data processing, and data storage facilities in several countries, which are owned by various parties (each with their own funding sources), and which are collectively operated by the ILT foundation under a joint scientific policy. The ILT resources have benefited from the following recent major funding sources: CNRS-INSU, Observatoire de Paris and Université d'Orléans, France; BMBF, MIWF-NRW, MPG, Germany; Science Foundation Ireland (SFI), Department of Business, Enterprise and Innovation (DBEI), Ireland; NWO, The Netherlands; The Science and Technology Facilities Council, UK; Ministry of Science and Higher Education, Poland[7].

Funding for the Sloan Digital Sky Survey V has been provided by the Alfred P. Sloan Foundation, the Heising-Simons Foundation, the National Science Foundation, and the Participating Institutions. SDSS acknowledges support and resources from the Center for High-Performance Computing at the University of Utah. SDSS telescopes are located at Apache Point Observatory, funded by the Astrophysical Research Consortium and operated by New Mexico State University, and at Las Campanas Observatory, operated by the Carnegie Institution for Science.

The Pan-STARRS1 Surveys (PS1) and the PS1 public science archive have been made possible through contributions by the Institute for Astronomy, the University of Hawaii, the Pan-STARRS Project Office, the Max-Planck Society and its participating institutes, the Max Planck Institute for Astronomy, Heidelberg and the Max Planck Institute for Extraterrestrial Physics, Garching, The Johns Hopkins University, Durham University, the University of Edinburgh, the Queen's University Belfast, the Harvard-Smithsonian Center for Astrophysics, the Las Cumbres Observatory Global Telescope Network Incorporated, the National Central University of Taiwan, the Space Telescope Science Institute, the National Aeronautics and Space Administration under Grant No. NNX08AR22G issued through the Planetary Science Division of the NASA Science Mission Directorate, the National Science Foundation Grant No. AST-1238877, the University of Maryland, Eotvos Lorand University (ELTE), the Los Alamos National Laboratory, and the Gordon and Betty Moore Foundation.

\end{acknowledgements}

\section*{ORCID IDs}
M. Br\"uggen \orcidlink{0000-0002-3369-7735} \href{https://orcid.org/0000-0002-3369-7735}{https://orcid.org/0000-0002-3369-7735} \\
H. Andernach \orcidlink{0000-0003-4873-1681}
\href{https://orcid.org/0000-0003-4873-1681}{https://orcid.org/0000-0003-4873-1681}\\

%
\bibliographystyle{aa} 
\bibliography{marco.bib} 

\begin{thebibliography}{108}
\expandafter\ifx\csname natexlab\endcsname\relax\def\natexlab#1{#1}\fi

\bibitem[{{Ahumada} {et~al.}(2020){Ahumada}, {Allende Prieto}, {Almeida},
  {Anders}, {Anderson}, {Andrews}, {Anguiano}, {Arcodia}, {Armengaud},
  {Aubert}, {Avila}, {Avila-Reese}, {Badenes}, {Balland}, {Barger},
  {Barrera-Ballesteros}, {Basu}, {Bautista}, {Beaton}, {Beers}, {Benavides},
  {Bender}, {Bernardi}, {Bershady}, {Beutler}, {Bidin}, {Bird}, {Bizyaev},
  {Blanc}, {Blanton}, {Boquien}, {Borissova}, {Bovy}, {Brandt}, {Brinkmann},
  {Brownstein}, {Bundy}, {Bureau}, {Burgasser}, {Burtin}, {Cano-D{\'\i}az},
  {Capasso}, {Cappellari}, {Carrera}, {Chabanier}, {Chaplin}, {Chapman},
  {Cherinka}, {Chiappini}, {Doohyun Choi}, {Chojnowski}, {Chung}, {Clerc},
  {Coffey}, {Comerford}, {Comparat}, {da Costa}, {Cousinou}, {Covey}, {Crane},
  {Cunha}, {Ilha}, {Dai}, {Damsted}, {Darling}, {Davidson}, {Davies}, {Dawson},
  {De}, {de la Macorra}, {De Lee}, {Queiroz}, {Deconto Machado}, {de la Torre},
  {Dell'Agli}, {du Mas des Bourboux}, {Diamond-Stanic}, {Dillon}, {Donor},
  {Drory}, {Duckworth}, {Dwelly}, {Ebelke}, {Eftekharzadeh}, {Davis Eigenbrot},
  {Elsworth}, {Eracleous}, {Erfanianfar}, {Escoffier}, {Fan}, {Farr},
  {Fern{\'a}ndez-Trincado}, {Feuillet}, {Finoguenov}, {Fofie},
  {Fraser-McKelvie}, {Frinchaboy}, {Fromenteau}, {Fu}, {Galbany}, {Garcia},
  {Garc{\'\i}a-Hern{\'a}ndez}, {Garma Oehmichen}, {Ge}, {Geimba Maia},
  {Geisler}, {Gelfand}, {Goddy}, {Gonzalez-Perez}, {Grabowski}, {Green},
  {Grier}, {Guo}, {Guy}, {Harding}, {Hasselquist}, {Hawken}, {Hayes}, {Hearty},
  {Hekker}, {Hogg}, {Holtzman}, {Horta}, {Hou}, {Hsieh}, {Huber}, {Hunt}, {Ider
  Chitham}, {Imig}, {Jaber}, {Jimenez Angel}, {Johnson}, {Jones},
  {J{\"o}nsson}, {Jullo}, {Kim}, {Kinemuchi}, {Kirkpatrick}, {Kite}, {Klaene},
  {Kneib}, {Kollmeier}, {Kong}, {Kounkel}, {Krishnarao}, {Lacerna}, {Lan},
  {Lane}, {Law}, {Le Goff}, {Leung}, {Lewis}, {Li}, {Lian}, {Lin}, {Long},
  {Longa-Pe{\~n}a}, {Lundgren}, {Lyke}, {Mackereth}, {MacLeod}, {Majewski},
  {Manchado}, {Maraston}, {Martini}, {Masseron}, {Masters}, {Mathur},
  {McDermid}, {Merloni}, {Merrifield}, {M{\'e}sz{\'a}ros}, {Miglio}, {Minniti},
  {Minsley}, {Miyaji}, {Mohammad}, {Mosser}, {Mueller}, {Muna},
  {Mu{\~n}oz-Guti{\'e}rrez}, {Myers}, {Nadathur}, {Nair}, {Nandra}, {Correa do
  Nascimento}, {Nevin}, {Newman}, {Nidever}, {Nitschelm}, {Noterdaeme},
  {O'Connell}, {Olmstead}, {Oravetz}, {Oravetz}, {Osorio}, {Pace}, {Padilla},
  {Palanque-Delabrouille}, {Palicio}, {Pan}, {Pan}, {Parker}, {Paviot},
  {Peirani}, {Ram{\'r}ez}, {Penny}, {Percival}, {Perez-Fournon},
  {P{\'e}rez-R{\`a}fols}, {Petitjean}, {Pieri}, {Pinsonneault}, {Poovelil},
  {Povick}, {Prakash}, {Price-Whelan}, {Raddick}, {Raichoor}, {Ray}, {Rembold},
  {Rezaie}, {Riffel}, {Riffel}, {Rix}, {Robin}, {Roman-Lopes},
  {Rom{\'a}n-Z{\'u}{\~n}iga}, {Rose}, {Ross}, {Rossi}, {Rowlands}, {Rubin},
  {Salvato}, {S{\'a}nchez}, {S{\'a}nchez-Menguiano}, {S{\'a}nchez-Gallego},
  {Sayres}, {Schaefer}, {Schiavon}, {Schimoia}, {Schlafly}, {Schlegel},
  {Schneider}, {Schultheis}, {Schwope}, {Seo}, {Serenelli}, {Shafieloo},
  {Shamsi}, {Shao}, {Shen}, {Shetrone}, {Shirley}, {Silva Aguirre}, {Simon},
  {Skrutskie}, {Slosar}, {Smethurst}, {Sobeck}, {Sodi}, {Souto}, {Stark},
  {Stassun}, {Steinmetz}, {Stello}, {Stermer}, {Storchi-Bergmann},
  {Streblyanska}, {Stringfellow}, {Stutz}, {Su{\'a}rez}, {Sun},
  {Taghizadeh-Popp}, {Talbot}, {Tayar}, {Thakar}, {Theriault}, {Thomas},
  {Thomas}, {Tinker}, {Tojeiro}, {Toledo}, {Tremonti}, {Troup}, {Tuttle},
  {Unda-Sanzana}, {Valentini}, {Vargas-Gonz{\'a}lez}, {Vargas-Maga{\~n}a},
  {V{\'a}zquez-Mata}, {Vivek}, {Wake}, {Wang}, {Weaver}, {Weijmans}, {Wild},
  {Wilson}, {Wilson}, {Wolthuis}, {Wood-Vasey}, {Yan}, {Yang}, {Y{\`e}che},
  {Zamora}, {Zarrouk}, {Zasowski}, {Zhang}, {Zhao}, {Zhao}, {Zheng}, {Zheng},
  {Zhu}, \& {Zou}}]{Ahumada2020}
{Ahumada}, R., {Allende Prieto}, C., {Almeida}, A., {et~al.} 2020, \apjs, 249,
  3

\bibitem[{{Alam} {et~al.}(2015){Alam}, {Albareti}, {Allende Prieto}, {Anders},
  {Anderson}, {Anderton}, {Andrews}, {Armengaud}, {Aubourg}, {Bailey}, {Basu},
  {Bautista}, {Beaton}, {Beers}, {Bender}, {Berlind}, {Beutler}, {Bhardwaj},
  {Bird}, {Bizyaev}, {Blake}, {Blanton}, {Blomqvist}, {Bochanski}, {Bolton},
  {Bovy}, {Shelden Bradley}, {Brandt}, {Brauer}, {Brinkmann}, {Brown},
  {Brownstein}, {Burden}, {Burtin}, {Busca}, {Cai}, {Capozzi}, {Carnero
  Rosell}, {Carr}, {Carrera}, {Chambers}, {Chaplin}, {Chen}, {Chiappini},
  {Chojnowski}, {Chuang}, {Clerc}, {Comparat}, {Covey}, {Croft}, {Cuesta},
  {Cunha}, {da Costa}, {Da Rio}, {Davenport}, {Dawson}, {De Lee}, {Delubac},
  {Deshpande}, {Dhital}, {Dutra-Ferreira}, {Dwelly}, {Ealet}, {Ebelke},
  {Edmondson}, {Eisenstein}, {Ellsworth}, {Elsworth}, {Epstein}, {Eracleous},
  {Escoffier}, {Esposito}, {Evans}, {Fan}, {Fern{\'a}ndez-Alvar}, {Feuillet},
  {Filiz Ak}, {Finley}, {Finoguenov}, {Flaherty}, {Fleming}, {Font-Ribera},
  {Foster}, {Frinchaboy}, {Galbraith-Frew}, {Garc{\'\i}a},
  {Garc{\'\i}a-Hern{\'a}ndez}, {Garc{\'\i}a P{\'e}rez}, {Gaulme}, {Ge},
  {G{\'e}nova-Santos}, {Georgakakis}, {Ghezzi}, {Gillespie}, {Girardi},
  {Goddard}, {Gontcho}, {Gonz{\'a}lez Hern{\'a}ndez}, {Grebel}, {Green},
  {Grieb}, {Grieves}, {Gunn}, {Guo}, {Harding}, {Hasselquist}, {Hawley},
  {Hayden}, {Hearty}, {Hekker}, {Ho}, {Hogg}, {Holley-Bockelmann}, {Holtzman},
  {Honscheid}, {Huber}, {Huehnerhoff}, {Ivans}, {Jiang}, {Johnson},
  {Kinemuchi}, {Kirkby}, {Kitaura}, {Klaene}, {Knapp}, {Kneib}, {Koenig},
  {Lam}, {Lan}, {Lang}, {Laurent}, {Le Goff}, {Leauthaud}, {Lee}, {Lee},
  {Licquia}, {Liu}, {Long}, {L{\'o}pez-Corredoira}, {Lorenzo-Oliveira},
  {Lucatello}, {Lundgren}, {Lupton}, {Mack}, {Mahadevan}, {Maia}, {Majewski},
  {Malanushenko}, {Malanushenko}, {Manchado}, {Manera}, {Mao}, {Maraston},
  {Marchwinski}, {Margala}, {Martell}, {Martig}, {Masters}, {Mathur},
  {McBride}, {McGehee}, {McGreer}, {McMahon}, {M{\'e}nard}, {Menzel},
  {Merloni}, {M{\'e}sz{\'a}ros}, {Miller}, {Miralda-Escud{\'e}}, {Miyatake},
  {Montero-Dorta}, {More}, {Morganson}, {Morice-Atkinson}, {Morrison},
  {Mosser}, {Muna}, {Myers}, {Nandra}, {Newman}, {Neyrinck}, {Nguyen},
  {Nichol}, {Nidever}, {Noterdaeme}, {Nuza}, {O'Connell}, {O'Connell},
  {O'Connell}, {Ogando}, {Olmstead}, {Oravetz}, {Oravetz}, {Osumi}, {Owen},
  {Padgett}, {Padmanabhan}, {Paegert}, {Palanque-Delabrouille}, {Pan},
  {Parejko}, {P{\^a}ris}, {Park}, {Pattarakijwanich}, {Pellejero-Ibanez},
  {Pepper}, {Percival}, {P{\'e}rez-Fournon}, {P{\'e}rez-R{\`a}fols},
  {Petitjean}, {Pieri}, {Pinsonneault}, {Porto de Mello}, {Prada}, {Prakash},
  {Price-Whelan}, {Protopapas}, {Raddick}, {Rahman}, {Reid}, {Rich}, {Rix},
  {Robin}, {Rockosi}, {Rodrigues}, {Rodr{\'\i}guez-Torres}, {Roe}, {Ross},
  {Ross}, {Rossi}, {Ruan}, {Rubi{\~n}o-Mart{\'\i}n}, {Rykoff},
  {Salazar-Albornoz}, {Salvato}, {Samushia}, {S{\'a}nchez}, {Santiago},
  {Sayres}, {Schiavon}, {Schlegel}, {Schmidt}, {Schneider}, {Schultheis},
  {Schwope}, {Sc{\'o}ccola}, {Scott}, {Sellgren}, {Seo}, {Serenelli}, {Shane},
  {Shen}, {Shetrone}, {Shu}, {Silva Aguirre}, {Sivarani}, {Skrutskie},
  {Slosar}, {Smith}, {Sobreira}, {Souto}, {Stassun}, {Steinmetz}, {Stello},
  {Strauss}, {Streblyanska}, {Suzuki}, {Swanson}, {Tan}, {Tayar}, {Terrien},
  {Thakar}, {Thomas}, {Thomas}, {Thompson}, {Tinker}, {Tojeiro}, {Troup},
  {Vargas-Maga{\~n}a}, {Vazquez}, {Verde}, {Viel}, {Vogt}, {Wake}, {Wang},
  {Weaver}, {Weinberg}, {Weiner}, {White}, {Wilson}, {Wisniewski},
  {Wood-Vasey}, {Ye`che}, {York}, {Zakamska}, {Zamora}, {Zasowski}, {Zehavi},
  {Zhao}, {Zheng}, {Zhou}, {Zhou}, {Zou}, \& {Zhu}}]{2015ApJS..219...12A}
{Alam}, S., {Albareti}, F.~D., {Allende Prieto}, C., {et~al.} 2015, \apjs, 219,
  12

\bibitem[{{Andernach} {et~al.}(2012){Andernach}, {Jim{\'e}nez Andrade},
  {Maldonado S{\'a}nchez}, \& {V{\'a}squez B{\'a}ez}}]{2012sngi.confP...1A}
{Andernach}, H., {Jim{\'e}nez Andrade}, E.~F., {Maldonado S{\'a}nchez}, R.~F.,
  \& {V{\'a}squez B{\'a}ez}, I.~R. 2012, in Science from the Next Generation
  Imaging and Spectroscopic Surveys, P1

\bibitem[{{Andernach} {et~al.}(2021){Andernach}, {Jim{\'e}nez-Andrade}, \&
  {Willis}}]{2021Galax...9...99A}
{Andernach}, H., {Jim{\'e}nez-Andrade}, E.~F., \& {Willis}, A.~G. 2021,
  Galaxies, 9, 99

\bibitem[{{Arshakian} \& {Longair}(2000)}]{Arshakian2000}
{Arshakian}, T.~G. \& {Longair}, M.~S. 2000, \mnras, 311, 846

\bibitem[{{Baldwin}(1982)}]{Baldwin1982}
{Baldwin}, J.~E. 1982, in Extragalactic Radio Sources, ed. D.~S. {Heeschen} \&
  C.~M. {Wade}, Vol.~97, 21--24

\bibitem[{{Banfield} {et~al.}(2015){Banfield}, {Wong}, {Willett}, {Norris},
  {Rudnick}, {Shabala}, {Simmons}, {Snyder}, {Garon}, {Seymour}, {Middelberg},
  {Andernach}, {Lintott}, {Jacob}, {Kapi{\'n}ska}, {Mao}, {Masters}, {Jarvis},
  {Schawinski}, {Paget}, {Simpson}, {Kl{\"o}ckner}, {Bamford}, {Burchell},
  {Chow}, {Cotter}, {Fortson}, {Heywood}, {Jones}, {Kaviraj},
  {L{\'o}pez-S{\'a}nchez}, {Maksym}, {Polsterer}, {Borden}, {Hollow}, \&
  {Whyte}}]{2015MNRAS.453.2326B}
{Banfield}, J.~K., {Wong}, O.~I., {Willett}, K.~W., {et~al.} 2015, \mnras, 453,
  2326

\bibitem[{{Barthel} {et~al.}(1985){Barthel}, {Schilizzi}, {Miley}, {J\"agers},
  \& {Strom}}]{Barthel1985}
{Barthel}, P.~D., {Schilizzi}, R.~T., {Miley}, G.~K., {J\"agers}, W.~J., \&
  {Strom}, R.~G. 1985, \aap, 148, 243

\bibitem[{{Bassani} {et~al.}(2021){Bassani}, {Ursini}, {Malizia}, {Bruni},
  {Panessa}, {Masetti}, {Saviane}, {Monaco}, {Venturi}, {Dallacasa}, {Bazzano},
  \& {Ubertini}}]{2021MNRAS.500.3111B}
{Bassani}, L., {Ursini}, F., {Malizia}, A., {et~al.} 2021, \mnras, 500, 3111

\bibitem[{{Beck} {et~al.}(2022){Beck}, {Dodds}, \&
  {Szapudi}}]{2022MNRAS.515.4711B}
{Beck}, R., {Dodds}, S.~C., \& {Szapudi}, I. 2022, \mnras, 515, 4711

\bibitem[{{Beck} {et~al.}(2021){Beck}, {Szapudi}, {Flewelling}, {Holmberg},
  {Magnier}, \& {Chambers}}]{2021MNRAS.500.1633B}
{Beck}, R., {Szapudi}, I., {Flewelling}, H., {et~al.} 2021, \mnras, 500, 1633

\bibitem[{{Becker} {et~al.}(1995){Becker}, {White}, \& {Helfand}}]{FIRST1995}
{Becker}, R.~H., {White}, R.~L., \& {Helfand}, D.~J. 1995, \apj, 450, 559

\bibitem[{{Blundell} {et~al.}(1999){Blundell}, {Rawlings}, \&
  {Willott}}]{Blundell1999}
{Blundell}, K.~M., {Rawlings}, S., \& {Willott}, C.~J. 1999, \aj, 117, 677

\bibitem[{{Bondi} {et~al.}(2024){Bondi}, {Scaramella}, {Zamorani}, {Ciliegi},
  {Vitello}, {Arias}, {Best}, {Bonato}, {Botteon}, {Brienza}, {Brunetti},
  {Hardcastle}, {Magliocchetti}, {Massaro}, {Morabito}, {Pentericci},
  {Prandoni}, {R{\"o}ttgering}, {Shimwell}, {Tasse}, {van Weeren}, \&
  {White}}]{2024A&A...683A.179B}
{Bondi}, M., {Scaramella}, R., {Zamorani}, G., {et~al.} 2024, \aap, 683, A179

\bibitem[{{Bonnarel} {et~al.}(2000){Bonnarel}, {Fernique}, {Bienaym{\'e}},
  {Egret}, {Genova}, {Louys}, {Ochsenbein}, {Wenger}, \&
  {Bartlett}}]{2000A&AS..143...33B}
{Bonnarel}, F., {Fernique}, P., {Bienaym{\'e}}, O., {et~al.} 2000, \aaps, 143,
  33

\bibitem[{{Brescia} {et~al.}(2014){Brescia}, {Cavuoti}, {Longo}, \& {De
  Stefano}}]{Brescia2014}
{Brescia}, M., {Cavuoti}, S., {Longo}, G., \& {De Stefano}, V. 2014, \aap, 568,
  A126

\bibitem[{{Br{\"u}ggen} {et~al.}(2021){Br{\"u}ggen}, {Reiprich}, {Bulbul},
  {Koribalski}, {Andernach}, {Rudnick}, {Hoang}, {Wilber}, {Duchesne},
  {Veronica}, {Pacaud}, {Hopkins}, {Norris}, {Johnston-Hollitt}, {Brown},
  {Bonafede}, {Brunetti}, {Collier}, {Sanders}, {Vardoulaki}, {Venturi},
  {Kapinska}, \& {Marvil}}]{Bruggen2021}
{Br{\"u}ggen}, M., {Reiprich}, T.~H., {Bulbul}, E., {et~al.} 2021, \aap, 647,
  A3

\bibitem[{{Condon} {et~al.}(1998){Condon}, {Cotton}, {Greisen}, {Yin},
  {Perley}, {Taylor}, \& {Broderick}}]{1998AJ....115.1693C}
{Condon}, J.~J., {Cotton}, W.~D., {Greisen}, E.~W., {et~al.} 1998, \aj, 115,
  1693

\bibitem[{{Dabhade} {et~al.}(2024){Dabhade}, {Chavan}, {Saikia}, {Oei}, \&
  {R{\"o}ttgering}}]{Dabhade2024}
{Dabhade}, P., {Chavan}, K., {Saikia}, D.~J., {Oei}, M. S.~S.~L., \&
  {R{\"o}ttgering}, H. J.~A. 2024, arXiv e-prints, arXiv:2408.13607

\bibitem[{{Dabhade} {et~al.}(2017){Dabhade}, {Gaikwad}, {Bagchi},
  {Pandey-Pommier}, {Sankhyayan}, \& {Raychaudhury}}]{Dabhade2017}
{Dabhade}, P., {Gaikwad}, M., {Bagchi}, J., {et~al.} 2017, \mnras, 469, 2886

\bibitem[{{Dabhade} {et~al.}(2020{\natexlab{a}}){Dabhade}, {Mahato}, {Bagchi},
  {Saikia}, {Combes}, {Sankhyayan}, {R{\"o}ttgering}, {Ho}, {Gaikwad},
  {Raychaudhury}, {Vaidya}, \& {Guiderdoni}}]{Dabhade2020a}
{Dabhade}, P., {Mahato}, M., {Bagchi}, J., {et~al.} 2020{\natexlab{a}}, \aap,
  642, A153

\bibitem[{{Dabhade} {et~al.}(2020{\natexlab{b}}){Dabhade}, {R{\"o}ttgering},
  {Bagchi}, {Shimwell}, {Hardcastle}, {Sankhyayan}, {Morganti}, {Jamrozy},
  {Shulevski}, \& {Duncan}}]{2020A&A...635A...5D}
{Dabhade}, P., {R{\"o}ttgering}, H.~J.~A., {Bagchi}, J., {et~al.}
  2020{\natexlab{b}}, \aap, 635, A5

\bibitem[{{Dabhade} {et~al.}(2023){Dabhade}, {Saikia}, \&
  {Mahato}}]{2023JApA...44...13D}
{Dabhade}, P., {Saikia}, D.~J., \& {Mahato}, M. 2023, Journal of Astrophysics
  and Astronomy, 44, 13

\bibitem[{{de la Rosa Vald{\'e}s} \& {Andernach}(2019)}]{2019arXiv190809988D}
{de la Rosa Vald{\'e}s}, P.~A. \& {Andernach}, H. 2019, arXiv e-prints,
  arXiv:1908.09988

\bibitem[{{Debski} {et~al.}(2025){Debski}, {Zeimann}, {Hill}, {Schneider},
  {Morabito}, {Dalton}, {Jarvis}, {Mentuch Cooper}, {Ciardullo}, {Gawiser}, \&
  {Jurlin}}]{2025ApJ...978..101D}
{Debski}, M.~H., {Zeimann}, G.~R., {Hill}, G.~J., {et~al.} 2025, \apj, 978, 101

\bibitem[{{Delhaize} {et~al.}(2021){Delhaize}, {Heywood}, {Prescott}, {Jarvis},
  {Delvecchio}, {Whittam}, {White}, {Hardcastle}, {Hale}, {Afonso}, {Ao},
  {Brienza}, {Br{\"u}ggen}, {Collier}, {Daddi}, {Glowacki}, {Maddox},
  {Morabito}, {Prandoni}, {Randriamanakoto}, {Sekhar}, {An}, {Adams}, {Blyth},
  {Bowler}, {Leeuw}, {Marchetti}, {Randriamampandry}, {Thorat}, {Seymour},
  {Smirnov}, {Taylor}, {Tasse}, \& {Vaccari}}]{2021MNRAS.501.3833D}
{Delhaize}, J., {Heywood}, I., {Prescott}, M., {et~al.} 2021, \mnras, 501, 3833

\bibitem[{{DESI Collaboration} {et~al.}(2025){DESI Collaboration},
  {Abdul-Karim}, {Adame}, \& {Aguado}}]{2025arXiv250314745D}
{DESI Collaboration}, {Abdul-Karim}, M., {Adame}, A.~G., \& {Aguado}, D. 2025,
  arXiv e-prints, arXiv:2503.14745

\bibitem[{{Duchesne} {et~al.}(2023){Duchesne}, {Thomson}, {Pritchard}, {Lenc},
  {Moss}, {McConnell}, {Wieringa}, {Whiting}, {Wang}, {Wang}, {Rose}, {Raja},
  {Murphy}, {Leung}, {Huynh}, {Hotan}, {Hodgson}, \&
  {Heald}}]{2023PASA...40...34D}
{Duchesne}, S.~W., {Thomson}, A.~J.~M., {Pritchard}, J., {et~al.} 2023, \pasa,
  40, e034

\bibitem[{{Duncan}(2022)}]{Duncan2022}
{Duncan}, K.~J. 2022, \mnras, 512, 3662

\bibitem[{{Edler} {et~al.}(2023){Edler}, {de Gasperin}, {Shimwell},
  {Hardcastle}, {Boselli}, {Heesen}, {McCall}, {Bomans}, {Br{\"u}ggen},
  {Bulbul}, {Chy{\.z}y}, {Ignesti}, {Merloni}, {Pacaud}, {Reiprich}, {Roberts},
  {Rottgering}, \& {van Weeren}}]{2023A&A...676A..24E}
{Edler}, H.~W., {de Gasperin}, F., {Shimwell}, T.~W., {et~al.} 2023, \aap, 676,
  A24

\bibitem[{{Eilek} \& {Shore}(1989)}]{Eilek1989}
{Eilek}, J.~A. \& {Shore}, S.~N. 1989, \apj, 342, 187

\bibitem[{{Fanaroff} \& {Riley}(1974)}]{1974MNRAS.167P..31F}
{Fanaroff}, B.~L. \& {Riley}, J.~M. 1974, \mnras, 167, 31P

\bibitem[{{Flewelling} {et~al.}(2020){Flewelling}, {Magnier}, {Chambers},
  {Heasley}, {Holmberg}, {Huber}, {Sweeney}, {Waters}, {Calamida}, {Casertano},
  {Chen}, {Farrow}, {Hasinger}, {Henderson}, {Long}, {Metcalfe}, {Narayan},
  {Nieto-Santisteban}, {Norberg}, {Rest}, {Saglia}, {Szalay}, {Thakar},
  {Tonry}, {Valenti}, {Werner}, {White}, {Denneau}, {Draper}, {Hodapp},
  {Jedicke}, {Kaiser}, {Kudritzki}, {Price}, {Wainscoat}, {Chastel}, {McLean},
  {Postman}, \& {Shiao}}]{2020ApJS..251....7F}
{Flewelling}, H.~A., {Magnier}, E.~A., {Chambers}, K.~C., {et~al.} 2020, \apjs,
  251, 7

\bibitem[{{Gordon} {et~al.}(2021){Gordon}, {Boyce}, {O'Dea}, {Rudnick},
  {Andernach}, {Vantyghem}, {Baum}, {Bui}, {Dionyssiou}, {Safi-Harb}, \&
  {Sander}}]{2021ApJS..255...30G}
{Gordon}, Y.~A., {Boyce}, M.~M., {O'Dea}, C.~P., {et~al.} 2021, \apjs, 255, 30

\bibitem[{{G{\"u}rkan} {et~al.}(2022){G{\"u}rkan}, {Prandoni}, {O'Brien},
  {Raja}, {Marchetti}, {Vaccari}, {Driver}, {Taylor}, {Franzen}, {Brown},
  {Shabala}, {Andernach}, {Hopkins}, {Norris}, {Leahy}, {Bilicki},
  {Farajollahi}, {Galvin}, {Heald}, {Koribalski}, {An}, \&
  {Warhurst}}]{Gurkan2022}
{G{\"u}rkan}, G., {Prandoni}, I., {O'Brien}, A., {et~al.} 2022, \mnras, 512,
  6104

\bibitem[{{Hale} {et~al.}(2021){Hale}, {McConnell}, {Thomson}, {Lenc}, {Heald},
  {Hotan}, {Leung}, {Moss}, {Murphy}, {Pritchard}, {Sadler}, {Stewart}, \&
  {Whiting}}]{2021PASA...38...58H}
{Hale}, C.~L., {McConnell}, D., {Thomson}, A.~J.~M., {et~al.} 2021, \pasa, 38,
  e058

\bibitem[{{Hao} {et~al.}(2010){Hao}, {McKay}, {Koester}, {Rykoff}, {Rozo},
  {Annis}, {Wechsler}, {Evrard}, {Siegel}, {Becker}, {Busha}, {Gerdes},
  {Johnston}, \& {Sheldon}}]{2010ApJS..191..254H}
{Hao}, J., {McKay}, T.~A., {Koester}, B.~P., {et~al.} 2010, \apjs, 191, 254

\bibitem[{{Hardcastle}(2018)}]{Hardcastle2018}
{Hardcastle}, M.~J. 2018, \mnras, 475, 2768

\bibitem[{{Hardcastle} {et~al.}(2023){Hardcastle}, {Horton}, {Williams},
  {Duncan}, {Alegre}, {Barkus}, {Croston}, {Dickinson}, {Osinga},
  {R{\"o}ttgering}, {Sabater}, {Shimwell}, {Smith}, {Best}, {Botteon},
  {Br{\"u}ggen}, {Drabent}, {de Gasperin}, {G{\"u}rkan}, {Hajduk}, {Hale},
  {Hoeft}, {Jamrozy}, {Kunert-Bajraszewska}, {Kondapally}, {Magliocchetti},
  {Mahatma}, {Mostert}, {O'Sullivan}, {Pajdosz-{\'S}mierciak}, {Petley},
  {Pierce}, {Prandoni}, {Schwarz}, {Shulewski}, {Siewert}, {Stott}, {Tang},
  {Vaccari}, {Zheng}, {Bailey}, {Desbled}, {Goyal}, {Gonano}, {Hanset},
  {Kurtz}, {Lim}, {Mielle}, {Molloy}, {Roth}, {Terentev}, \&
  {Torres}}]{Hardcastle2023}
{Hardcastle}, M.~J., {Horton}, M.~A., {Williams}, W.~L., {et~al.} 2023, \aap,
  678, A151

\bibitem[{{Hardcastle} {et~al.}(2025){Hardcastle}, {Pierce}, {Duncan},
  {G{\"u}rkan}, {Gong}, {Horton}, {Mingo}, {R{\"o}ttgering}, \&
  {Smith}}]{2025arXiv250409303H}
{Hardcastle}, M.~J., {Pierce}, J.~C.~S., {Duncan}, K.~J., {et~al.} 2025, arXiv
  e-prints, arXiv:2504.09303

\bibitem[{{Helfand} {et~al.}(2015){Helfand}, {White}, \&
  {Becker}}]{2015ApJ...801...26H}
{Helfand}, D.~J., {White}, R.~L., \& {Becker}, R.~H. 2015, \apj, 801, 26

\bibitem[{{Hurley-Walker} {et~al.}(2017){Hurley-Walker}, {Callingham},
  {Hancock}, {Franzen}, {Hindson}, {Kapi{\'n}ska}, {Morgan}, {Offringa},
  {Wayth}, {Wu}, {Zheng}, {Murphy}, {Bell}, {Dwarakanath}, {For}, {Gaensler},
  {Johnston-Hollitt}, {Lenc}, {Procopio}, {Staveley-Smith}, {Ekers}, {Bowman},
  {Briggs}, {Cappallo}, {Deshpande}, {Greenhill}, {Hazelton}, {Kaplan},
  {Lonsdale}, {McWhirter}, {Mitchell}, {Morales}, {Morgan}, {Oberoi}, {Ord},
  {Prabu}, {Shankar}, {Srivani}, {Subrahmanyan}, {Tingay}, {Webster},
  {Williams}, \& {Williams}}]{2017MNRAS.464.1146H}
{Hurley-Walker}, N., {Callingham}, J.~R., {Hancock}, P.~J., {et~al.} 2017,
  \mnras, 464, 1146

\bibitem[{{Hurley-Walker} {et~al.}(2022){Hurley-Walker}, {Galvin}, {Duchesne},
  {Zhang}, {Morgan}, {Hancock}, {An}, {Franzen}, {Heald}, {Ross}, {Vernstrom},
  {Anderson}, {Gaensler}, {Johnston-Hollitt}, {Kaplan}, {Riseley}, {Tingay}, \&
  {Walker}}]{2022PASA...39...35H}
{Hurley-Walker}, N., {Galvin}, T.~J., {Duchesne}, S.~W., {et~al.} 2022, \pasa,
  39, e035

\bibitem[{{Intema} {et~al.}(2017){Intema}, {Jagannathan}, {Mooley}, \&
  {Frail}}]{2017A&A...598A..78I}
{Intema}, H.~T., {Jagannathan}, P., {Mooley}, K.~P., \& {Frail}, D.~A. 2017,
  \aap, 598, A78

\bibitem[{{Ishwara-Chandra} \&
  {Saikia}(1999{\natexlab{a}})}]{1999MNRAS.309..100I}
{Ishwara-Chandra}, C.~H. \& {Saikia}, D.~J. 1999{\natexlab{a}}, \mnras, 309,
  100

\bibitem[{{Ishwara-Chandra} \&
  {Saikia}(1999{\natexlab{b}})}]{Ishwara-Chandra1999}
{Ishwara-Chandra}, C.~H. \& {Saikia}, D.~J. 1999{\natexlab{b}}, \mnras, 309,
  100

\bibitem[{{Ishwara-Chandra} \& {Saikia}(1999{\natexlab{c}})}]{Ishwara1999}
{Ishwara-Chandra}, C.~H. \& {Saikia}, D.~J. 1999{\natexlab{c}}, \mnras, 309,
  100

\bibitem[{{Itoh} {et~al.}(2020){Itoh}, {Utsumi}, {Inoue}, {Ohta}, {Doi},
  {Morokuma}, {Kawabata}, \& {Tanaka}}]{2020ApJ...901....3I}
{Itoh}, R., {Utsumi}, Y., {Inoue}, Y., {et~al.} 2020, \apj, 901, 3

\bibitem[{{Knowles} {et~al.}(2022){Knowles}, {Cotton}, {Rudnick}, {Camilo},
  {Goedhart}, {Deane}, {Ramatsoku}, {Bietenholz}, {Br{\"u}ggen}, {Button},
  {Chen}, {Chibueze}, {Clarke}, {de Gasperin}, {Ianjamasimanana}, {J{\'o}zsa},
  {Hilton}, {Kesebonye}, {Kolokythas}, {Kraan-Korteweg}, {Lawrie}, {Lochner},
  {Loubser}, {Marchegiani}, {Mhlahlo}, {Moodley}, {Murphy}, {Namumba},
  {Oozeer}, {Parekh}, {Pillay}, {Passmoor}, {Ramaila}, {Ranchod},
  {Retana-Montenegro}, {Sebokolodi}, {Sikhosana}, {Smirnov}, {Thorat},
  {Venturi}, {Abbott}, {Adam}, {Adams}, {Aldera}, {Bauermeister}, {Bennett},
  {Bode}, {Botha}, {Botha}, {Brederode}, {Buchner}, {Burger}, {Cheetham}, {de
  Villiers}, {Dikgale-Mahlakoana}, {du Toit}, {Esterhuyse}, {Fadana},
  {Fanaroff}, {Fataar}, {Foley}, {Fourie}, {Frank}, {Gamatham}, {Gatsi},
  {Geyer}, {Gouws}, {Gumede}, {Heywood}, {Hlakola}, {Hokwana}, {Hoosen},
  {Horn}, {Horrell}, {Hugo}, {Isaacson}, {Jonas}, {Jordaan}, {Joubert},
  {Julie}, {Kapp}, {Kasper}, {Kenyon}, {Kotz{\'e}}, {Kotze}, {Kriek}, {Kriel},
  {Krishnan}, {Kusel}, {Legodi}, {Lehmensiek}, {Liebenberg}, {Lord}, {Lunsky},
  {Madisa}, {Magnus}, {Main}, {Makhaba}, {Makhathini}, {Malan}, {Manley},
  {Marais}, {Maree}, {Martens}, {Mauch}, {McAlpine}, {Merry}, {Millenaar},
  {Mokone}, {Monama}, {Mphego}, {New}, {Ngcebetsha}, {Ngoasheng}, {Ockards},
  {Otto}, {Patel}, {Peens-Hough}, {Perkins}, {Ramanujam}, {Ramudzuli},
  {Ratcliffe}, {Renil}, {Robyntjies}, {Rust}, {Salie}, {Sambu}, {Schollar},
  {Schwardt}, {Schwartz}, {Serylak}, {Siebrits}, {Sirothia}, {Slabber},
  {Sofeya}, {Taljaard}, {Tasse}, {Tiplady}, {Toruvanda}, {Twum}, {van Balla},
  {van der Byl}, {van der Merwe}, {van Dyk}, {Van Tonder}, {Van Wyk}, {Venter},
  {Venter}, {Welz}, {Williams}, \& {Xaia}}]{2022A&A...657A..56K}
{Knowles}, K., {Cotton}, W.~D., {Rudnick}, L., {et~al.} 2022, \aap, 657, A56

\bibitem[{{Komberg} \& {Pashchenko}(2009)}]{Komberg2009}
{Komberg}, B.~V. \& {Pashchenko}, I.~N. 2009, Astronomy Reports, 53, 1086

\bibitem[{{Koribalski}(2025)}]{2025arXiv250407314K}
{Koribalski}, B.~S. 2025, arXiv e-prints, arXiv:2504.07314

\bibitem[{{Krogager} {et~al.}(2018){Krogager}, {Gupta}, {Noterdaeme}, {Ranjan},
  {Fynbo}, {Srianand}, {Petitjean}, {Combes}, \&
  {Mahabal}}]{2018ApJS..235...10K}
{Krogager}, J.~K., {Gupta}, N., {Noterdaeme}, P., {et~al.} 2018, \apjs, 235, 10

\bibitem[{{Kronberg} {et~al.}(1986){Kronberg}, {Wielebinski}, \&
  {Graham}}]{1986A&A...169...63K}
{Kronberg}, P.~P., {Wielebinski}, R., \& {Graham}, D.~A. 1986, \aap, 169, 63

\bibitem[{{Ku{\'z}micz} \& {Jamrozy}(2012)}]{Kuzmicz2012}
{Ku{\'z}micz}, A. \& {Jamrozy}, M. 2012, \mnras, 426, 851

\bibitem[{{Ku{\'z}micz} \& {Jamrozy}(2021)}]{Kuzmicz2021}
{Ku{\'z}micz}, A. \& {Jamrozy}, M. 2021, \apjs, 253, 25

\bibitem[{{Ku{\'z}micz} {et~al.}(2018){Ku{\'z}micz}, {Jamrozy}, {Bronarska},
  {Janda-Boczar}, \& {Saikia}}]{Kuzmicz2018}
{Ku{\'z}micz}, A., {Jamrozy}, M., {Bronarska}, K., {Janda-Boczar}, K., \&
  {Saikia}, D.~J. 2018, \apjs, 238, 9

\bibitem[{{Lacy} {et~al.}(2020){Lacy}, {Baum}, {Chandler}, {Chatterjee},
  {Clarke}, {Deustua}, {English}, {Farnes}, {Gaensler}, {Gugliucci},
  {Hallinan}, {Kent}, {Kimball}, {Law}, {Lazio}, {Marvil}, {Mao}, {Medlin},
  {Mooley}, {Murphy}, {Myers}, {Osten}, {Richards}, {Rosolowsky}, {Rudnick},
  {Schinzel}, {Sivakoff}, {Sjouwerman}, {Taylor}, {White}, {Wrobel},
  {Andernach}, {Beasley}, {Berger}, {Bhatnager}, {Birkinshaw}, {Bower},
  {Brandt}, {Brown}, {Burke-Spolaor}, {Butler}, {Comerford}, {Demorest}, {Fu},
  {Giacintucci}, {Golap}, {G{\"u}th}, {Hales}, {Hiriart}, {Hodge}, {Horesh},
  {Ivezi{\'c}}, {Jarvis}, {Kamble}, {Kassim}, {Liu}, {Loinard}, {Lyons},
  {Masters}, {Mezcua}, {Moellenbrock}, {Mroczkowski}, {Nyland}, {O'Dea},
  {O'Sullivan}, {Peters}, {Radford}, {Rao}, {Robnett}, {Salcido}, {Shen},
  {Sobotka}, {Witz}, {Vaccari}, {van Weeren}, {Vargas}, {Williams}, \&
  {Yoon}}]{2020PASP..132c5001L}
{Lacy}, M., {Baum}, S.~A., {Chandler}, C.~J., {et~al.} 2020, \pasp, 132, 035001

\bibitem[{{Lan} \& {Prochaska}(2021)}]{Lan2021}
{Lan}, T.-W. \& {Prochaska}, X.~J. 2021, \mnras, 502, 5104

\bibitem[{{Lara} {et~al.}(2004){Lara}, {Giovannini}, {Cotton}, {Feretti},
  {Marcaide}, {M{\'a}rquez}, \& {Venturi}}]{Lara2004}
{Lara}, L., {Giovannini}, G., {Cotton}, W.~D., {et~al.} 2004, \aap, 421, 899

\bibitem[{{Lara} {et~al.}(2001){Lara}, {M{\'a}rquez}, {Cotton}, {Feretti},
  {Giovannini}, {Marcaide}, \& {Venturi}}]{Lara2001}
{Lara}, L., {M{\'a}rquez}, I., {Cotton}, W.~D., {et~al.} 2001, \aap, 378, 826

\bibitem[{{Lynch} {et~al.}(2021){Lynch}, {Galvin}, {Line}, {Jordan}, {Trott},
  {Chege}, {McKinley}, {Johnston-Hollitt}, \& {Tingay}}]{2021PASA...38...57L}
{Lynch}, C.~R., {Galvin}, T.~J., {Line}, J.~L.~B., {et~al.} 2021, \pasa, 38,
  e057

\bibitem[{{Machalski} {et~al.}(2004){Machalski}, {Chyzy}, \&
  {Jamrozy}}]{Machalski2004}
{Machalski}, J., {Chyzy}, K.~T., \& {Jamrozy}, M. 2004, \actaa, 54, 391

\bibitem[{{Machalski} {et~al.}(2001){Machalski}, {Jamrozy}, \&
  {Zola}}]{Machalski2001}
{Machalski}, J., {Jamrozy}, M., \& {Zola}, S. 2001, \aap, 371, 445

\bibitem[{{Mahato} {et~al.}(2022){Mahato}, {Dabhade}, {Saikia}, {Combes},
  {Bagchi}, {Ho}, \& {Raychaudhury}}]{Mahato2022}
{Mahato}, M., {Dabhade}, P., {Saikia}, D.~J., {et~al.} 2022, \aap, 660, A59

\bibitem[{{Mauch} {et~al.}(2003){Mauch}, {Murphy}, {Buttery}, {Curran},
  {Hunstead}, {Piestrzynski}, {Robertson}, \& {Sadler}}]{SUMSS2003}
{Mauch}, T., {Murphy}, T., {Buttery}, H.~J., {et~al.} 2003, \mnras, 342, 1117

\bibitem[{{McConnell} {et~al.}(2020){McConnell}, {Hale}, {Lenc}, {Banfield},
  {Heald}, {Hotan}, {Leung}, {Moss}, {Murphy}, {O'Brien}, {Pritchard}, {Raja},
  {Sadler}, {Stewart}, {Thomson}, {Whiting}, {Allison}, {Amy}, {Anderson},
  {Ball}, {Bannister}, {Bell}, {Bock}, {Bolton}, {Bunton}, {Chippendale},
  {Collier}, {Cooray}, {Cornwell}, {Diamond}, {Edwards}, {Gupta}, {Hayman},
  {Heywood}, {Jackson}, {Koribalski}, {Lee-Waddell}, {McClure-Griffiths}, {Ng},
  {Norris}, {Phillips}, {Reynolds}, {Roxby}, {Schinckel}, {Shields},
  {Tremblay}, {Tzioumis}, {Voronkov}, \& {Westmeier}}]{2020PASA...37...48M}
{McConnell}, D., {Hale}, C.~L., {Lenc}, E., {et~al.} 2020, \pasa, 37, e048

\bibitem[{{Mostert} {et~al.}(2023){Mostert}, {Morganti}, {Brienza}, {Duncan},
  {Oei}, {R{\"o}ttgering}, {Alegre}, {Hardcastle}, \&
  {Jurlin}}]{2023A&A...674A.208M}
{Mostert}, R. I.~J., {Morganti}, R., {Brienza}, M., {et~al.} 2023, \aap, 674,
  A208

\bibitem[{{Mostert} {et~al.}(2024){Mostert}, {Oei}, {Barkus}, {Alegre},
  {Hardcastle}, {Duncan}, {R{\"o}ttgering}, {van Weeren}, \&
  {Horton}}]{2024A&A...691A.185M}
{Mostert}, R.~I.~J., {Oei}, M.~S.~S.~L., {Barkus}, B., {et~al.} 2024, \aap,
  691, A185

\bibitem[{{Neronov} {et~al.}(2025){Neronov}, {Vazza}, {Brandenburg}, \&
  {Caprini}}]{2025A&A...696L...8N}
{Neronov}, A., {Vazza}, F., {Brandenburg}, A., \& {Caprini}, C. 2025, \aap,
  696, L8

\bibitem[{{Norris} {et~al.}(2021){Norris}, {Marvil}, {Collier}, {Kapi{\'n}ska},
  {O'Brien}, {Rudnick}, {Andernach}, {Asorey}, {Brown}, {Br{\"u}ggen},
  {Crawford}, {English}, {Rahman}, {Filipovi{\'c}}, {Gordon}, {G{\"u}rkan},
  {Hale}, {Hopkins}, {Huynh}, {HyeongHan}, {James Jee}, {Koribalski}, {Lenc},
  {Luken}, {Parkinson}, {Prandoni}, {Raja}, {Reiprich}, {Riseley}, {Shabala},
  {Sheil}, {Vernstrom}, {Whiting}, {Allison}, {Anderson}, {Ball}, {Bell},
  {Bunton}, {Galvin}, {Gupta}, {Hotan}, {Jacka}, {Macgregor}, {Mahony}, {Maio},
  {Moss}, {Pandey-Pommier}, \& {Voronkov}}]{2021PASA...38...46N}
{Norris}, R.~P., {Marvil}, J., {Collier}, J.~D., {et~al.} 2021, \pasa, 38, e046

\bibitem[{{Oei} {et~al.}(2024){Oei}, {Hardcastle}, {Timmerman}, {Gast},
  {Botteon}, {Rodriguez}, {Stern}, {Calistro Rivera}, {van Weeren},
  {R{\"o}ttgering}, {Intema}, {de Gasperin}, \&
  {Djorgovski}}]{2024Natur.633..537O}
{Oei}, M. S.~S.~L., {Hardcastle}, M.~J., {Timmerman}, R., {et~al.} 2024, \nat,
  633, 537

\bibitem[{{Oei} {et~al.}(2023{\natexlab{a}}){Oei}, {van Weeren}, {Gast},
  {Botteon}, {Hardcastle}, {Dabhade}, {Shimwell}, {R{\"o}ttgering}, \&
  {Drabent}}]{2023A&A...672A.163O}
{Oei}, M. S.~S.~L., {van Weeren}, R.~J., {Gast}, A. R.~D.~J.~G.~I.~B., {et~al.}
  2023{\natexlab{a}}, \aap, 672, A163

\bibitem[{{Oei} {et~al.}(2022){Oei}, {van Weeren}, {Hardcastle}, {Botteon},
  {Shimwell}, {Dabhade}, {Gast}, {R{\"o}ttgering}, {Br{\"u}ggen}, {Tasse},
  {Williams}, \& {Shulevski}}]{2022A&A...660A...2O}
{Oei}, M. S.~S.~L., {van Weeren}, R.~J., {Hardcastle}, M.~J., {et~al.} 2022,
  \aap, 660, A2

\bibitem[{{Oei} {et~al.}(2023{\natexlab{b}}){Oei}, {van Weeren}, {Hardcastle},
  {Vazza}, {Shimwell}, {Leclercq}, {Br{\"u}ggen}, \&
  {R{\"o}ttgering}}]{2023MNRAS.518..240O}
{Oei}, M. S.~S.~L., {van Weeren}, R.~J., {Hardcastle}, M.~J., {et~al.}
  2023{\natexlab{b}}, \mnras, 518, 240

\bibitem[{{Offringa} {et~al.}(2016){Offringa}, {Trott}, {Hurley-Walker},
  {Johnston-Hollitt}, {McKinley}, {Barry}, {Beardsley}, {Bowman}, {Briggs},
  {Carroll}, {Dillon}, {Ewall-Wice}, {Feng}, {Gaensler}, {Greenhill},
  {Hazelton}, {Hewitt}, {Jacobs}, {Kim}, {Kittiwisit}, {Lenc}, {Line}, {Loeb},
  {Mitchell}, {Morales}, {Neben}, {Paul}, {Pindor}, {Pober}, {Procopio},
  {Riding}, {Sethi}, {Shankar}, {Subrahmanyan}, {Sullivan}, {Tegmark},
  {Thyagarajan}, {Tingay}, {Wayth}, {Webster}, \&
  {Wyithe}}]{2016MNRAS.458.1057O}
{Offringa}, A.~R., {Trott}, C.~M., {Hurley-Walker}, N., {et~al.} 2016, \mnras,
  458, 1057

\bibitem[{{Oguri} {et~al.}(2018){Oguri}, {Lin}, {Lin}, {Nishizawa}, {More},
  {More}, {Hsieh}, {Medezinski}, {Miyatake}, {Jian}, {Lin}, {Takada}, {Okabe},
  {Speagle}, {Coupon}, {Leauthaud}, {Lupton}, {Miyazaki}, {Price}, {Tanaka},
  {Chiu}, {Komiyama}, {Okura}, {Tanaka}, \& {Usuda}}]{2018PASJ...70S..20O}
{Oguri}, M., {Lin}, Y.-T., {Lin}, S.-C., {et~al.} 2018, \pasj, 70, S20

\bibitem[{{O'Sullivan} {et~al.}(2023){O'Sullivan}, {Shimwell}, {Hardcastle},
  {Tasse}, {Heald}, {Carretti}, {Br{\"u}ggen}, {Vacca}, {Sobey}, {Van Eck},
  {Horellou}, {Beck}, {Bilicki}, {Bourke}, {Botteon}, {Croston}, {Drabent},
  {Duncan}, {Heesen}, {Ideguchi}, {Kirwan}, {Lawlor}, {Mingo},
  {Nikiel-Wroczy{\'n}ski}, {Piotrowska}, {Scaife}, \& {van
  Weeren}}]{2023MNRAS.519.5723O}
{O'Sullivan}, S.~P., {Shimwell}, T.~W., {Hardcastle}, M.~J., {et~al.} 2023,
  \mnras, 519, 5723

\bibitem[{Perucho(2019)}]{galaxies7030070}
Perucho, M. 2019, Galaxies, 7, 70

\bibitem[{{Quici} {et~al.}(2025){Quici}, {Turner}, {Seymour}, \&
  {Hurley-Walker}}]{2025MNRAS.537..343Q}
{Quici}, B., {Turner}, R.~J., {Seymour}, N., \& {Hurley-Walker}, N. 2025,
  \mnras, 537, 343

\bibitem[{{Rengelink} {et~al.}(1997){Rengelink}, {Tang}, {de Bruyn}, {Miley},
  {Bremer}, {Roettgering}, \& {Bremer}}]{WENSS1997}
{Rengelink}, R.~B., {Tang}, Y., {de Bruyn}, A.~G., {et~al.} 1997, \aaps, 124,
  259

\bibitem[{{Sankhyayan} \& {Dabhade}(2024)}]{2024A&A...687L...8S}
{Sankhyayan}, S. \& {Dabhade}, P. 2024, \aap, 687, L8

\bibitem[{{Saripalli} {et~al.}(2005){Saripalli}, {Hunstead}, {Subrahmanyan}, \&
  {Boyce}}]{Saripalli2005}
{Saripalli}, L., {Hunstead}, R.~W., {Subrahmanyan}, R., \& {Boyce}, E. 2005,
  \aj, 130, 896

\bibitem[{{Schoenmakers} {et~al.}(2001){Schoenmakers}, {de Bruyn},
  {R{\"o}ttgering}, \& {van der Laan}}]{2001A&A...374..861S}
{Schoenmakers}, A.~P., {de Bruyn}, A.~G., {R{\"o}ttgering}, H.~J.~A., \& {van
  der Laan}, H. 2001, \aap, 374, 861

\bibitem[{{Schoenmakers} {et~al.}(2000){Schoenmakers}, {Mack}, {de Bruyn},
  {R{\"o}ttgering}, {Klein}, \& {van der Laan}}]{2000A&AS..146..293S}
{Schoenmakers}, A.~P., {Mack}, K.~H., {de Bruyn}, A.~G., {et~al.} 2000, \aaps,
  146, 293

\bibitem[{{Sejake} {et~al.}(2023){Sejake}, {White}, {Heywood}, {Thorat},
  {Bester}, {Makhathini}, \& {Fanaroff}}]{2023MNRAS.518.4290S}
{Sejake}, P.~K., {White}, S.~V., {Heywood}, I., {et~al.} 2023, \mnras, 518,
  4290

\bibitem[{{Shimwell} {et~al.}(2022){Shimwell}, {Hardcastle}, {Tasse}, {Best},
  {R{\"o}ttgering}, {Williams}, {Botteon}, {Drabent}, {Mechev}, {Shulevski},
  {van Weeren}, {Bester}, {Br{\"u}ggen}, {Brunetti}, {Callingham}, {Chy{\.z}y},
  {Conway}, {Dijkema}, {Duncan}, {de Gasperin}, {Hale}, {Haverkorn}, {Hugo},
  {Jackson}, {Mevius}, {Miley}, {Morabito}, {Morganti}, {Offringa}, {Oonk},
  {Rafferty}, {Sabater}, {Smith}, {Schwarz}, {Smirnov}, {O'Sullivan},
  {Vedantham}, {White}, {Albert}, {Alegre}, {Asabere}, {Bacon}, {Bonafede},
  {Bonnassieux}, {Brienza}, {Bilicki}, {Bonato}, {Calistro Rivera}, {Cassano},
  {Cochrane}, {Croston}, {Cuciti}, {Dallacasa}, {Danezi}, {Dettmar}, {Di
  Gennaro}, {Edler}, {En{\ss}lin}, {Emig}, {Franzen}, {Garc{\'\i}a-Vergara},
  {Grange}, {G{\"u}rkan}, {Hajduk}, {Heald}, {Heesen}, {Hoang}, {Hoeft},
  {Horellou}, {Iacobelli}, {Jamrozy}, {Jeli{\'c}}, {Kondapally}, {Kukreti},
  {Kunert-Bajraszewska}, {Magliocchetti}, {Mahatma}, {Ma{\l}ek}, {Mandal},
  {Massaro}, {Meyer-Zhao}, {Mingo}, {Mostert}, {Nair}, {Nakoneczny},
  {Nikiel-Wroczy{\'n}ski}, {Orr{\'u}}, {Pajdosz-{\'S}mierciak}, {Pasini},
  {Prandoni}, {van Piggelen}, {Rajpurohit}, {Retana-Montenegro}, {Riseley},
  {Rowlinson}, {Saxena}, {Schrijvers}, {Sweijen}, {Siewert}, {Timmerman},
  {Vaccari}, {Vink}, {West}, {Wo{\l}owska}, {Zhang}, \&
  {Zheng}}]{2022A&A...659A...1S}
{Shimwell}, T.~W., {Hardcastle}, M.~J., {Tasse}, C., {et~al.} 2022, \aap, 659,
  A1

\bibitem[{{Shimwell} {et~al.}(2019){Shimwell}, {Tasse}, {Hardcastle}, {Mechev},
  {Williams}, {Best}, {R{\"o}ttgering}, {Callingham}, {Dijkema}, {de Gasperin},
  {Hoang}, {Hugo}, {Mirmont}, {Oonk}, {Prandoni}, {Rafferty}, {Sabater},
  {Smirnov}, {van Weeren}, {White}, {Atemkeng}, {Bester}, {Bonnassieux},
  {Br{\"u}ggen}, {Brunetti}, {Chy{\.z}y}, {Cochrane}, {Conway}, {Croston},
  {Danezi}, {Duncan}, {Haverkorn}, {Heald}, {Iacobelli}, {Intema}, {Jackson},
  {Jamrozy}, {Jarvis}, {Lakhoo}, {Mevius}, {Miley}, {Morabito}, {Morganti},
  {Nisbet}, {Orr{\'u}}, {Perkins}, {Pizzo}, {Schrijvers}, {Smith}, {Vermeulen},
  {Wise}, {Alegre}, {Bacon}, {van Bemmel}, {Beswick}, {Bonafede}, {Botteon},
  {Bourke}, {Brienza}, {Calistro Rivera}, {Cassano}, {Clarke}, {Conselice},
  {Dettmar}, {Drabent}, {Dumba}, {Emig}, {En{\ss}lin}, {Ferrari}, {Garrett},
  {G{\'e}nova-Santos}, {Goyal}, {G{\"u}rkan}, {Hale}, {Harwood}, {Heesen},
  {Hoeft}, {Horellou}, {Jackson}, {Kokotanekov}, {Kondapally},
  {Kunert-Bajraszewska}, {Mahatma}, {Mahony}, {Mandal}, {McKean}, {Merloni},
  {Mingo}, {Miskolczi}, {Mooney}, {Nikiel-Wroczy{\'n}ski}, {O'Sullivan},
  {Quinn}, {Reich}, {Roskowi{\'n}ski}, {Rowlinson}, {Savini}, {Saxena},
  {Schwarz}, {Shulevski}, {Sridhar}, {Stacey}, {Urquhart}, {van der Wiel},
  {Varenius}, {Webster}, \& {Wilber}}]{Shimwell2019}
{Shimwell}, T.~W., {Tasse}, C., {Hardcastle}, M.~J., {et~al.} 2019, \aap, 622,
  A1

\bibitem[{{Simonte} {et~al.}(2023){Simonte}, {Andernach}, {Br{\"u}ggen},
  {Best}, \& {Osinga}}]{Simonte2023}
{Simonte}, M., {Andernach}, H., {Br{\"u}ggen}, M., {Best}, P.~N., \& {Osinga},
  E. 2023, \aap, 672, A178

\bibitem[{{Simonte} {et~al.}(2024){Simonte}, {Andernach}, {Br{\"u}ggen},
  {Miley}, \& {Barthel}}]{Simonte2024}
{Simonte}, M., {Andernach}, H., {Br{\"u}ggen}, M., {Miley}, G.~K., \&
  {Barthel}, P. 2024, \aap, 686, A21

\bibitem[{{Simonte} {et~al.}(2022){Simonte}, {Andernach}, {Br{\"u}ggen},
  {Schwarz}, {Prandoni}, \& {Willis}}]{Simonte2022}
{Simonte}, M., {Andernach}, H., {Br{\"u}ggen}, M., {et~al.} 2022, \mnras, 515,
  2032

\bibitem[{{Szabo} {et~al.}(2011){Szabo}, {Pierpaoli}, {Dong}, {Pipino}, \&
  {Gunn}}]{2011ApJ...736...21S}
{Szabo}, T., {Pierpaoli}, E., {Dong}, F., {Pipino}, A., \& {Gunn}, J. 2011,
  \apj, 736, 21

\bibitem[{{Thongkham} {et~al.}(2024){Thongkham}, {Gonzalez}, {Brodwin},
  {Trudeau}, {Eisenhardt}, {Stanford}, {Moravec}, {Connor}, {Stern}, {Spivey},
  \& {Garcia}}]{2024ApJ...976..186T}
{Thongkham}, K., {Gonzalez}, A.~H., {Brodwin}, M., {et~al.} 2024, \apj, 976,
  186

\bibitem[{{Turner} \& {Shabala}(2015)}]{Turner2015}
{Turner}, R.~J. \& {Shabala}, S.~S. 2015, \apj, 806, 59

\bibitem[{{Turner} {et~al.}(2023){Turner}, {Yates-Jones}, {Shabala}, {Quici},
  \& {Stewart}}]{Turner2023}
{Turner}, R.~J., {Yates-Jones}, P.~M., {Shabala}, S.~S., {Quici}, B., \&
  {Stewart}, G. S.~C. 2023, \mnras, 518, 945

\bibitem[{{van Haarlem} {et~al.}(2013){van Haarlem}, {Wise}, {Gunst}, {Heald},
  {McKean}, {Hessels}, {de Bruyn}, {Nijboer}, {Swinbank}, {Fallows},
  {Brentjens}, {Nelles}, {Beck}, {Falcke}, {Fender}, {H{\"o}randel},
  {Koopmans}, {Mann}, {Miley}, {R{\"o}ttgering}, {Stappers}, {Wijers},
  {Zaroubi}, {van den Akker}, {Alexov}, {Anderson}, {Anderson}, {van Ardenne},
  {Arts}, {Asgekar}, {Avruch}, {Batejat}, {B{\"a}hren}, {Bell}, {Bell}, {van
  Bemmel}, {Bennema}, {Bentum}, {Bernardi}, {Best}, {B{\^\i}rzan}, {Bonafede},
  {Boonstra}, {Braun}, {Bregman}, {Breitling}, {van de Brink}, {Broderick},
  {Broekema}, {Brouw}, {Br{\"u}ggen}, {Butcher}, {van Cappellen}, {Ciardi},
  {Coenen}, {Conway}, {Coolen}, {Corstanje}, {Damstra}, {Davies}, {Deller},
  {Dettmar}, {van Diepen}, {Dijkstra}, {Donker}, {Doorduin}, {Dromer}, {Drost},
  {van Duin}, {Eisl{\"o}ffel}, {van Enst}, {Ferrari}, {Frieswijk}, {Gankema},
  {Garrett}, {de Gasperin}, {Gerbers}, {de Geus}, {Grie{\ss}meier}, {Grit},
  {Gruppen}, {Hamaker}, {Hassall}, {Hoeft}, {Holties}, {Horneffer}, {van der
  Horst}, {van Houwelingen}, {Huijgen}, {Iacobelli}, {Intema}, {Jackson},
  {Jelic}, {de Jong}, {Juette}, {Kant}, {Karastergiou}, {Koers}, {Kollen},
  {Kondratiev}, {Kooistra}, {Koopman}, {Koster}, {Kuniyoshi}, {Kramer},
  {Kuper}, {Lambropoulos}, {Law}, {van Leeuwen}, {Lemaitre}, {Loose}, {Maat},
  {Macario}, {Markoff}, {Masters}, {McFadden}, {McKay-Bukowski}, {Meijering},
  {Meulman}, {Mevius}, {Middelberg}, {Millenaar}, {Miller-Jones}, {Mohan},
  {Mol}, {Morawietz}, {Morganti}, {Mulcahy}, {Mulder}, {Munk}, {Nieuwenhuis},
  {van Nieuwpoort}, {Noordam}, {Norden}, {Noutsos}, {Offringa}, {Olofsson},
  {Omar}, {Orr{\'u}}, {Overeem}, {Paas}, {Pandey-Pommier}, {Pandey}, {Pizzo},
  {Polatidis}, {Rafferty}, {Rawlings}, {Reich}, {de Reijer}, {Reitsma},
  {Renting}, {Riemers}, {Rol}, {Romein}, {Roosjen}, {Ruiter}, {Scaife}, {van
  der Schaaf}, {Scheers}, {Schellart}, {Schoenmakers}, {Schoonderbeek},
  {Serylak}, {Shulevski}, {Sluman}, {Smirnov}, {Sobey}, {Spreeuw}, {Steinmetz},
  {Sterks}, {Stiepel}, {Stuurwold}, {Tagger}, {Tang}, {Tasse}, {Thomas},
  {Thoudam}, {Toribio}, {van der Tol}, {Usov}, {van Veelen}, {van der Veen},
  {ter Veen}, {Verbiest}, {Vermeulen}, {Vermaas}, {Vocks}, {Vogt}, {de Vos},
  {van der Wal}, {van Weeren}, {Weggemans}, {Weltevrede}, {White}, {Wijnholds},
  {Wilhelmsson}, {Wucknitz}, {Yatawatta}, {Zarka}, {Zensus}, \& {van
  Zwieten}}]{LOFAR2013}
{van Haarlem}, M.~P., {Wise}, M.~W., {Gunst}, A.~W., {et~al.} 2013, \aap, 556,
  A2

\bibitem[{{Villarreal Hern{\'a}ndez} \&
  {Andernach}(2018)}]{2018arXiv180807178V}
{Villarreal Hern{\'a}ndez}, A.~C. \& {Andernach}, H. 2018, arXiv e-prints,
  arXiv:1808.07178

\bibitem[{{Wagenveld} {et~al.}(2024){Wagenveld}, {Kl{\"o}ckner}, {Gupta},
  {Sekhar}, {Jagannathan}, {Deka}, {Jose}, {Balashev}, {Borgaonkar},
  {Chatterjee}, {Combes}, {Emig}, {Gaunekar}, {Hilton}, {J{\'o}zsa}, {Klutse},
  {Knowles}, {Krogager}, {Momjian}, {Muller}, \&
  {Sikhosana}}]{2024A&A...690A.163W}
{Wagenveld}, J.~D., {Kl{\"o}ckner}, H.~R., {Gupta}, N., {et~al.} 2024, \aap,
  690, A163

\bibitem[{{Wen} \& {Han}(2015)}]{2015ApJ...807..178W}
{Wen}, Z.~L. \& {Han}, J.~L. 2015, \apj, 807, 178

\bibitem[{{Wen} \& {Han}(2024)}]{2024ApJS..272...39W}
{Wen}, Z.~L. \& {Han}, J.~L. 2024, \apjs, 272, 39

\bibitem[{{White} {et~al.}(2019){White}, {Long}, {Becker}, {Blair}, {Helfand},
  \& {Winkler}}]{2019ApJS..241...37W}
{White}, R.~L., {Long}, K.~S., {Becker}, R.~H., {et~al.} 2019, \apjs, 241, 37

\bibitem[{{Willis} {et~al.}(1974){Willis}, {Strom}, \& {Wilson}}]{Willis1974}
{Willis}, A.~G., {Strom}, R.~G., \& {Wilson}, A.~S. 1974, \nat, 250, 625

\bibitem[{{Wong} {et~al.}(2025){Wong}, {Garon}, {Alger}, {Rudnick}, {Shabala},
  {Willett}, {Banfield}, {Andernach}, {Norris}, {Swan}, {Hardcastle},
  {Lintott}, {White}, {Seymour}, {Kapi{\'n}ska}, {Tang}, {Simmons}, \&
  {Schawinski}}]{2025MNRAS.536.3488W}
{Wong}, O.~I., {Garon}, A.~F., {Alger}, M.~J., {et~al.} 2025, \mnras, 536, 3488

\bibitem[{{Yang} {et~al.}(2021){Yang}, {Xu}, {He}, {Gu}, {Katsianis}, {Meng},
  {Shi}, {Zou}, {Zhang}, {Liu}, {Wang}, {Dong}, {Lu}, {Li}, {Chen}, {Wang},
  {Mo}, {Fu}, {Guo}, {Leauthaud}, {Luo}, {Zhang}, \&
  {Zu}}]{2021ApJ...909..143Y}
{Yang}, X., {Xu}, H., {He}, M., {et~al.} 2021, \apj, 909, 143

\bibitem[{{Yantovski-Barth} {et~al.}(2024){Yantovski-Barth}, {Newman}, {Dey},
  {Andrews}, {Eracleous}, {Golden-Marx}, \& {Zhou}}]{2024MNRAS.531.2285Y}
{Yantovski-Barth}, M.~J., {Newman}, J.~A., {Dey}, B., {et~al.} 2024, \mnras,
  531, 2285

\bibitem[{{Zhou} {et~al.}(2021){Zhou}, {Newman}, {Mao}, {Meisner}, {Moustakas},
  {Myers}, {Prakash}, {Zentner}, {Brooks}, {Duan}, {Landriau}, {Levi}, {Prada},
  \& {Tarle}}]{Zhou2021}
{Zhou}, R., {Newman}, J.~A., {Mao}, Y.-Y., {et~al.} 2021, \mnras, 501, 3309

\bibitem[{{Zhou} {et~al.}(2025){Zhou}, {Li}, {Zou}, {Gong}, {Deng}, {Chen},
  {Yu}, {He}, \& {Ding}}]{2025MNRAS.536.2260Z}
{Zhou}, X., {Li}, N., {Zou}, H., {et~al.} 2025, \mnras, 536, 2260

\bibitem[{{Zou} {et~al.}(2021){Zou}, {Gao}, {Xu}, {Zhou}, {Ma}, {Zhou},
  {Zhang}, {Nie}, {Wang}, \& {Xue}}]{2021ApJS..253...56Z}
{Zou}, H., {Gao}, J., {Xu}, X., {et~al.} 2021, \apjs, 253, 56

\bibitem[{{Zou} {et~al.}(2022){Zou}, {Sui}, {Xue}, {Zhou}, {Ma}, {Zhou}, {Nie},
  {Zhang}, {Feng}, {Shen}, \& {Wang}}]{2022RAA....22f5001Z}
{Zou}, H., {Sui}, J., {Xue}, S., {et~al.} 2022, Research in Astronomy and
  Astrophysics, 22, 065001

\end{thebibliography}
%

%
%
%
%

\begin{appendix} 
\label{sec:bestfit}


\begin{table*}
\centering
\caption{Sample of 143 GRGs, sorted by increasing right ascension. Columns are:
(1) name of the giant radio galaxy from truncated sexagesimal RA,DEC (J2000) of the host galaxy;
(2,3) right ascension and declination (J2000) of the host galaxy in degrees;
(4) redshift of the host galaxy;
(5) redshift type of the host galaxy: $e$ for our own estimate, since no 
$z_{phot}$ is available in literature; $p$ for photometric, being the average of values found in various references listed in \citet{Simonte2024}, plus a few more recent ones
\citep{2022MNRAS.515.4711B,2024ApJS..272...39W,2025MNRAS.536.2260Z};
$s$ for spectroscopic, includung the recent DESI\,DR1 \citep{2025arXiv250314745D}; 
for 4 objects marked as 's*' redshifts were obtained by us in 2015 and 2016 with 
the 10-m GranTeCan (GTC) on Canary Islands;
(6) host type: G = galaxy, GP = galaxy pair, Q = quasar, QB = quasar with broad lines in SDSS spectrum, Qc = quasar candidate;
(7) largest angular size in arcmin;
(8) LLS in Mpc; (9) bending angle in degrees;
(10) armlength ratio of brighter-to-fainter lobe;
(11) survey used for flux integration: T = TGSS at 150\,MHz \citep{2017A&A...598A..78I},
L = LOFAR at 145\,MHz \citep{2022A&A...659A...1S,2023A&A...676A..24E},
R = RACS \citep{2020PASA...37...48M,2023PASA...40...34D}  at 888 and 1367\,MHz;
A = ASKAP at 944\,MHz \citep{2021PASA...38...46N} or SWAG-X (V.\ Moss, priv.\ comm) at 888 MHz;
N = NVSS at 1.4\,GHz \citep{1998AJ....115.1693C};
(12) total flux at the survey frequency;
(13) decimal log of power at 145\,MHz extrapolated from the survey frequency with $\alpha=-0.8$;
(14) code for provenance: P = from published literature (mainly 
\citet{Kuzmicz2018,2021Galax...9...99A,2023A&A...672A.163O,Hardcastle2023,Simonte2024,2024A&A...691A.185M},\cite{2025arXiv250407314K}),
N = new discovery by us, mostly from NVSS or LoTSS, or otherwise found by us in:
MA = MeerKAT MALS DR2 \citep{2024A&A...690A.163W} at 1.27~MHz,
MW = MWA: EoR images \citep{2016MNRAS.458.1057O,2021PASA...38...57L}, 
GLEAM \citep{2017MNRAS.464.1146H} or GLEAM-X DR1 \citep{2022PASA...39...35H},
A = ASKAP-EMU (Evolutionary Map of the Universe, \cite{2021PASA...38...46N} or other ASKAP project, 
ED = LOFAR Euclid Deep Field North \citep{2024A&A...683A.179B},
R = Radio Galaxy Zoo \citep{2015MNRAS.453.2326B,2025MNRAS.536.3488W}, 
V = VICTORIA \citep{2023A&A...676A..24E},
n = mentioned in literature, but not as GRG.
(15) environment flag: b = brightest cluster galaxy, m = cluster member galaxy, i =  isolated galaxy
(see sect.~\ref{sec:clusters} for references).}

\footnotesize{
    \begin{tabular}{lclccrcccccccl}
\hline
(1) & (2) ~~~~~~~~ (3) & (4) & (5) & (6) & (7) & (8) & (9) & (10) & (11) & (12) & (13) & (14) & (15) \\
GRG~Jname & RA ~~~~~~ Dec & ~~~$z$ & ztyp & host & LAS & LLS & BA & ALR & surv & $S_{\rm{\nu},surv}$ &  $\log P_{145}$ & prov & env \\
       & $^{\circ}$ ~~~~~~~~~~~~ $^{\circ}$ &  & &  & $'$~~~ & Mpc & $^{\circ}$ &
       &  & mJy & W/Hz &  &  \\ \hline

J0006$+$2402     & 001.5395 $+$24.0393 & 0.78   & p & G  &  6.77 & 3.02 &  1.0 & 0.94 & L  &  107.5 & 26.42 & P  & i  \\
J0011$+$3100     & 002.9694 $+$31.0068 & 0.68   & p & G  &  7.12 & 3.02 &  9.4 & 0.71 & L  &  223.0 & 26.60 & P  &    \\
J0019$-$0408     & 004.9126 $-$04.1492 & 0.75   & p & G  &  7.0  & 3.08 &  8.1 & 0.58 & A  &  110.0 & 27.04 & MW &    \\
J0030$-$0312     & 007.5070 $-$03.2095 & 0.543  & s*& G  &  8.0  & 3.06 &  3.3 & 0.75 & R  &  114.1 & 26.71 & N  &    \\
J0039$-$1928     & 009.8106 $-$19.4799 & 0.78   & p & G  &  6.9  & 3.08 &  0.6 & 0.96 & R  &   63.0 & 26.82 & MA &    \\
J0041$+$2912     & 010.3823 $+$29.2049 & 1.2482 & s & QB &  6.34 & 3.17 &  0.0 & 0.55 & L  &   98.3 & 26.86 & P  &    \\
J0044$-$3804     & 011.0582 $-$38.0765 & 1.0    & p & G  &  6.9  & 3.32 &  2.0 & 2.19 & A  &   25.8 & 26.70 & A  & i  \\
J0045$+$2847     & 011.2868 $+$28.7951 & 0.70   & p & G  &  7.7  & 3.30 &  0.0 & 0.78 & L  &  108.9 & 26.32 & P  &    \\
J0047$-$6206     & 011.9663 $-$62.1129 & 0.42   & p & G  & 10.53 & 3.49 &  3.0 & 0.50 & A  &  135.0 & 26.54 & A  & b  \\
J0048$+$0210     & 012.2001 $+$02.1676 & 0.3605 & s & G  & 11.3  & 3.41 &  3.8 & 0.83 & R  &   28.9 & 25.70 & P  & b  \\
J0054$-$4952     & 013.7436 $-$49.8739 & 0.47   & p & G  &  9.67 & 3.42 &  1.0 & 0.84 & R  &   27.6 & 25.95 & P  &    \\
J0059$-$2352     & 014.9781 $-$23.8816 & 0.727  & p & G  &   7.9 & 3.44 &  3.1 & 0.82 & A  &   16.4 & 26.18 & P & i   \\
J0101$+$5052$^a$ & 015.4744 $+$50.8830 & 1.6    & p & Qc &  6.15 & 3.13 &  4.7 & 1.04 & T  &  654.0 & 27.93 & N  &    \\
J0103$+$2305     & 015.8522 $+$23.0841 & 0.265  & s & G  & 24.1  & 5.90 &  3.8 & 0.62 & L  &  305.1 & 25.80 & n  &    \\
J0105$-$0808     & 016.3609 $-$08.1346 & 1.0    & p & Qc &  6.65 & 3.20 &  1.0 & 0.88 & A  &   40.2 & 26.90 & A  &    \\
J0107$-$2347     & 016.8392 $-$23.7928 & 0.30   & p & G  &  13.8 & 3.68 &  0.8 & 1.06 & A  &   89.0 & 26.03 & P &     \\
J0123$+$2936     & 020.9258 $+$29.6092 & 1.175  & s & Qc &  6.35 & 3.15 &  7.2 & 1.06 & L  &   87.5 & 26.75 & P  &    \\
J0124$+$1940     & 021.1689 $+$19.6678 & 0.44   & p & G  &  9.7  & 3.30 & 24.0 & 0.75 & L  &   80.7 & 25.72 & P  &    \\
J0134$+$3015$^b$ & 023.5443 $+$30.2605 & 0.54   & p & G  &  8.14 & 3.10 &  4.9 & 0.75 & L  &  337.0 & 26.54 & N  &    \\
J0153$+$5946     & 028.3051 $+$59.7810 & 0.3    & p & G  & 12.0  & 3.20 &  7.0 & 0.77 & T  &  645.0 & 26.24 & N  &    \\
J0153$-$2713     & 028.4932 $-$27.2272 & 0.2444 & s & G  & 16.5  & 3.80 &  3.5 & 0.94 & R  &   44.3 & 25.51 & N  & bm \\
J0156$-$6529     & 029.1281 $-$65.4917 & 0.64   & p & G  &  8.38 & 3.50 &  1.0 & 1.02 & A  &   37.9 & 26.42 & A  &    \\
J0156$-$0714     & 029.1806 $-$07.2345 & 1.18   & p & G  &  8.67 & 4.30 &  2.0 & 0.60 & MA &   13.8 & 26.70 & MA &    \\
J0157$-$6000     & 029.4660 $-$60.0161 & 0.837  & p & G  & 10.0  & 4.57 &  1.0 & 0.93 & A  &   35.0 & 26.66 & A  & i  \\
J0226$+$5927     & 036.6064 $+$59.4640 & 1.3    & p & Qc &  8.73 & 4.39 &  8.9 & 0.59 & T  & 2476.0 & 28.30 & P  &    \\
J0228$-$0207     & 037.1934 $-$02.1304 & 1.05   & p & G  &  8.49 & 4.13 &  0.0 & 0.51 & A  &   25.6 & 26.75 & A  &    \\
J0230$-$5230     & 037.5461 $-$52.5005 & 0.96   & p & G  &  6.51 & 3.10 &  3.0 & 0.88 & A  &   25.0 & 26.65 & A  &    \\
J0235$+$3104     & 038.9373 $+$31.0799 & 0.51   & p & G  &  9.5  & 3.52 & 17.5 & 1.82 & L  &  391.0 & 26.55 & P  &    \\
J0319$-$5340     & 049.8637 $-$53.6828 & 0.95   & p & G  &  6.9  & 3.27 &  4.7 & 0.70 & A  &  179.0 & 27.49 & A  &    \\
J0320$-$5330     & 050.1480 $-$53.5145 & 0.255  & p & G  & 17.7  & 4.22 &  4.5 & 1.47 & R  &  164.0 & 26.12 & N  &    \\
J0331$-$7713     & 052.9158 $-$77.2220 & 0.1456 & s & G  & 20.0  & 3.06 &  0.9 & 0.52 & R  &  557.2 & 26.11 & P  &    \\
J0352$-$2451     & 058.1742 $-$24.8619 & 0.35   & p & G  & 12.8  & 3.79 &  2.9 & 0.72 & MA &    7.8 & 25.23 & MA &    \\
J0509$+$0420     & 077.4607 $+$04.3392 & 0.23   & p & G  & 17.0  & 3.74 &  3.0 & 1.05 & R  &   62.8 & 25.60 & n  &    \\
J0648$+$7049$^c$ & 102.0173 $+$70.8276 & 0.75   & p & G  &  7.8  & 3.44 &      &      & L  &        &       & P  & i  \\
J0648$+$3917     & 102.1237 $+$39.2999 & 0.43   & p & G  &  8.94 & 3.01 &  0.0 & 0.69 & R  &   45.0 & 26.07 & N  &    \\
J0722$+$4242$^c$ & 110.6594 $+$42.7076 & 0.4290 & s & G  &  9.5  & 3.19 &  8.5 & 0.59 & N  &   24.0 & 25.96 & P  & b  \\
J0740$-$6647     & 115.2189 $-$66.7926 & 0.268  & p & G  & 12.7  & 3.13 &  3.0 & 1.05 & A  &   74.5 & 25.84 & A  & b  \\
J0742$-$6407     & 115.5709 $-$64.1196 & 0.43   & p & Qc & 10.08 & 3.39 &  5.5 & 1.09 & R  &  426.9 & 27.05 & N  &    \\
J0814$+$5224     & 123.5904 $+$52.4028 & 0.2467 & s & G  & 20.4  & 4.74 & 22.0 & 1.46 & L  &  413.2 & 25.90 & P  & b  \\
J0819$+$3235     & 124.9851 $+$32.5938 & 0.7535 & s & G  & 10.56 & 4.66 &  2.9 & 0.98 & L  &   70.9 & 26.20 & P  & i  \\
J0827$+$6628     & 126.9492 $+$66.4712 & 0.97   & p & G  &  7.84 & 3.74 &  2.5 & 0.85 & L  &  102.7 & 26.62 & P  &    \\
J0831$+$6711     & 127.9214 $+$67.1933 & 0.4274 & s & G  &  9.07 & 3.04 &  3.5 & 0.89 & L  &   62.9 & 25.58 & P  & b  \\
J0838$+$5327     & 129.5074 $+$53.4540 & 0.5445 & s & G  & 17.2  & 6.58 &  0.5 & 0.64 & L  &  117.0 & 26.09 & n  &    \\
J0839$+$0213     & 129.8990 $+$02.2324 & 0.7368 & s & G  &  9.08 & 3.97 &  5.2 & 1.02 & R  &   13.1 & 26.08 & A  & b  \\
J0841$+$5546     & 130.3626 $+$55.7742 & 0.7912 & s & G  &  6.9  & 3.10 &  3.6 & 0.68 & L  &  244.9 & 26.79 & P  & i  \\
J0843$+$0208$^a$ & 130.9196 $+$02.1398 & 0.88   & p & G  & 10.45 & 4.83 &  0.9 & 0.53 & R  &   24.5 & 26.53 & A  & i  \\
J0846$+$3747     & 131.6035 $+$37.7942 & 0.5476 & s & G  & 12.5  & 4.80 &  1.4 & 0.59 & L  &  192.3 & 26.31 & N  & b  \\
J0901$+$4420     & 135.3586 $+$44.3451 & 0.87   & p & G? & 11.7  & 5.40 &  0.0 & 1.47 & L  &   42.1 & 26.12 & N  &    \\
J0905$+$5630     & 136.3939 $+$56.5145 & 0.86   & p & G? &  6.6  & 3.05 &  4.0 & 0.95 & L  &   32.6 & 26.00 & P  &    \\
J0905$+$5526$^d$ & 136.4012 $+$55.4345 & 0.99   & p & G  &  7.26 & 3.49 &  1.5 & 0.38 & L  &   64.4 & 26.44 & n  &    \\
\label{tab:list}                
\end{tabular}}
\end{table*}

\begin{table*}


\centering
\footnotesize{
\begin{tabular}{lclccrcccrcccl}
J0907$+$6706     & 136.9662 $+$67.1070 & 0.6    & p & G  &  8.0  & 3.21 &  0.0 & 0.71 & L  &   58.5 & 25.89 & P  &    \\
J0911$+$5440     & 137.7929 $+$54.6736 & 0.1826 & s & G  & 18.0  & 3.30 & 32.0 & 0.47 & L  &   40.4 & 24.56 & N  & b  \\
J0923$+$0742     & 140.8455 $+$07.7037 & 0.55   & p & G  &  8.9  & 3.40 &  5.4 & 0.70 & R  &   81.1 & 26.57 & N  &    \\
J0928$+$2304$^c$ & 142.1122 $+$23.0800 & 0.67   & p & G  &  8.4  & 3.62 &  0.0 & 0.93 & R  &   17.0 & 26.10 & P  & b  \\
J0931$+$3204     & 142.9127 $+$32.0667 & 0.2257 & s & QB & 19.85 & 4.31 &  1.1 & 0.69 & L  &  268.1 & 25.58 & P  &    \\
J0939$+$0540     & 144.9669 $+$05.6772 & 0.84   & p & G  &  7.05 & 3.31 &  7.0 & 0.69 & R  &   99.0 & 27.09 & A  & i  \\
J0955$-$0744     & 148.7783 $-$07.7468 & 0.1567 & s & G  & 20.2  & 3.28 &  4.9 & 0.65 & R  &   28.0 & 24.90 & A  & b  \\
J1006$+$3454     & 151.5073 $+$34.9029 & 0.1005 & s & G  & 39.0  & 4.33 &  1.9 & 0.67 & L  & 17744. & 26.64 & P  &    \\
J1006$+$4834     & 151.7041 $+$48.5770 & 0.9    & p & G  &  6.95 & 3.25 & 12.0 & 1.40 & L  &   30.0 & 26.01 & P  &    \\
J1007$+$6847     & 151.9480 $+$68.7926 & 0.76   & p & G  &  9.65 & 4.27 &  7.4 & 0.79 & L  &   90.1 & 26.32 & N  &    \\
J1008$-$2652     & 152.2406 $-$26.8723 & 0.3    & p & G  & 12.0  & 3.20 &  0.0 & 1.29 & R  &   26.5 & 25.48 & MW &    \\
J1024$+$3818     & 156.1289 $+$38.3119 & 0.41   & p & G  & 10.3  & 3.37 &  2.0 & 1.13 & L  &   55.8 & 25.49 & P  &    \\
J1036$+$3732     & 159.0191 $+$37.5494 & 0.95   & p & G  &  7.06 & 3.35 &  3.0 & 1.43 & L  &   22.5 & 25.94 & P  &    \\
J1037$+$3129     & 159.3811 $+$31.4969 & 0.5229 & s & G  &  8.0  & 3.00 &  1.0 & 0.95 & L  &   80.9 & 25.89 & P  &    \\
J1050$+$5706     & 162.5251 $+$57.1100 & 1.0    & p & G  &  8.15 & 3.91 &  3.0 & 0.90 & L  &   24.7 & 26.03 & N  &    \\
J1101$-$1151     & 165.4437 $-$11.8649 & 0.66   & p & G  &  9.0  & 3.76 &  5.0 & 0.94 & MA &   20.2 & 26.28 & MA &    \\
J1102$+$1345     & 165.7198 $+$13.7545 & 0.671  & s*& G  &  7.3  & 3.07 &  1.0 & 0.75 & R  &  449.7 & 27.52 & R  & mi \\
J1108$+$2916     & 167.1436 $+$29.2753 & 0.8    & p & G  & 11.56 & 5.20 &  3.5 & 0.64 & L  &  286.5 & 26.87 & P  &    \\
J1113$+$2100     & 168.4967 $+$21.0165 & 0.7937 & s & Q  &  8.15 & 3.66 &  2.8 & 0.79 & R  &   30.3 & 26.52 & A  &    \\
J1116$-$4234     & 169.0210 $-$42.5804 & 0.26   & p & Qc & 13.0  & 3.10 &  1.0 & 1.57 & R  &   42.8 & 25.55 & MW &    \\
J1118$+$6433     & 169.5503 $+$64.5514 & 0.374  & p & G  & 12.3  & 3.80 &  4.0 & 0.52 & L  &   90.4 & 25.02 & P  & i  \\
J1120$+$3118     & 170.0495 $+$31.3144 & 0.49   & p & G  & 13.0  & 4.71 &  0.0 & 1.02 & L  &  100.5 & 25.92 & n  & i  \\
J1126$+$3025     & 171.6598 $+$30.4283 & 0.3049 & s & G  & 11.48 & 3.10 &  2.0 & 1.20 & L  &  456.0 & 26.11 & P  & b  \\
J1129$+$2733     & 172.3006 $+$27.5539 & 0.0732 & s & G  & 40.0  & 3.34 & 29.0 & 1.14 & N  &  157.0 & 25.10 & P  &    \\
J1150$+$4618     & 177.5346 $+$46.3108 & 1.0    & p & Qc &  7.05 & 3.39 &  1.0 & 0.98 & L  &  153.8 & 26.83 & N  &    \\
J1200$+$6523     & 180.1704 $+$65.3970 & 1.0    & p & Qc &  9.00 & 4.44 & 11.0 & 0.89 & L  &   99.0 & 26.64 & P  &    \\
J1202$+$0211     & 180.6776 $+$02.1860 & 0.88   & p & G  &  8.08 & 3.76 &  2.0 & 0.70 & A  &  145.0 & 27.32 & A  &    \\
J1210$+$4808     & 182.7070 $+$48.1354 & 0.85   & p & G  &  8.85 & 4.07 &  0.5 & 0.82 & L  &   78.4 & 26.37 & N  & i  \\
J1213$-$1534     & 183.3612 $-$15.5807 & 0.67   & p & Qc &  7.71 & 3.25 &  6.0 & 1.68 & R  &  150.2 & 27.04 & N  &    \\
J1218$+$3824     & 184.5653 $+$38.4021 & 0.8    & p & G  & 10.6  & 4.80 &  4.5 & 2.12 & L  &   47.8 & 26.09 & P  & i  \\
J1220$+$3837     & 185.1692 $+$38.6274 & 0.97   & p & G  &  6.7  & 3.20 &  1.4 & 1.14 & L  &   31.5 & 26.11 & N  &    \\
J1227$+$1437     & 186.9460 $+$14.6241 & 0.75   & p & G  &  9.95 & 4.38 &  2.6 & 1.07 & L  &   79.5 & 26.25 & V  & bi \\
J1229$+$5826     & 187.3728 $+$58.4442 & 0.77   & p & G  &  6.8  & 3.00 &  6.7 & 0.86 & L  &  160.2 & 26.58 & N  & i  \\
J1235$+$5317     & 188.7564 $+$53.2986 & 0.3448 & s & G  & 11.17 & 3.25 & 13.7 & 0.86 & L  &  753.6 & 26.75 & P  & b  \\
J1239$+$3609     & 189.7529 $+$36.1568 & 0.5935 & s & QB &  8.6  & 3.43 &  3.0 & 1.14 & L  &   61.8 & 25.90 & P  &    \\
J1245$+$4604     & 191.3971 $+$46.0795 & 0.85   & p & G  &  8.35 & 3.84 &  2.4 & 0.91 & L  &   59.6 & 26.25 & N  &    \\
J1301$-$0333     & 195.2897 $-$03.5517 & 0.7390 & s & Q  &  6.86 & 3.00 &  3.0 & 0.47 & A  &   78.2 & 26.88 & A  & i  \\
J1303$+$5404$^e$ & 195.8005 $+$54.0695 & 0.85   & p & G  &  8.93 & 4.10 &  6.0 & 0.99 & L  &   34.0 & 26.01 & N  & i  \\
J1305$-$3814     & 196.4313 $-$38.2456 & 0.3    & p & G  & 11.8  & 3.15 &  1.5 & 0.73 & R  &  132.0 & 26.18 & N  &    \\
J1318$+$2626     & 199.5976 $+$26.4397 & 0.6230 & s & QB & 10.72 & 4.37 &  0.1 & 0.78 & L  &  157.1 & 26.36 & P  &    \\
J1320$+$6022$^f$ & 200.1000 $+$60.3764 & 0.355  & p & G  & 12.5  & 3.74 & 16.0 & 0.77 & L  &  105.5 & 25.62 & P  &    \\
J1331$+$2934     & 202.7741 $+$29.5767 & 0.73   & p & G  &  7.65 & 3.33 &  6.0 & 1.62 & L  &  329.1 & 26.84 & P  & bi \\
J1344$+$2912     & 206.1514 $+$29.2110 & 0.82   & p & G  &  6.7  & 3.04 &  2.3 & 1.05 & L  &  107.1 & 26.47 & P  &    \\
J1411$+$6259     & 212.8681 $+$62.9986 & 1.1    & p & G  &  6.52 & 3.20 &  3.6 & 1.24 & L  &  181.7 & 27.00 & P  &    \\
J1416$+$4101     & 214.2086 $+$41.0307 & 0.7    & p & G  &  7.1  & 3.05 &  1.7 & 1.25 & L  &   80.3 & 26.18 & P  & i  \\
J1420$-$0545     & 215.0992 $-$05.7580 & 0.3067 & s & G  & 17.85 & 4.84 &  1.0 & 0.94 & R  &  138.7 & 26.22 & P  & mi \\
J1423$+$3529     & 215.9201 $+$35.4888 & 0.7771 & s & G  &  7.39 & 3.34 &  4.0 & 1.70 & L  &   31.7 & 25.89 & P  & i  \\
J1430$+$4104     & 217.5497 $+$41.0678 & 0.5868 & s & G  &  7.7  & 3.06 &  2.5 & 1.08 & L  &  197.7 & 26.40 & P  & i  \\
J1438$+$3355     & 219.5874 $+$33.9287 & 1.0    & p & Qc &  7.49 & 3.60 &  2.0 & 0.70 & L  &   25.0 & 26.04 & P  &    \\
J1450$+$6723     & 222.6387 $+$67.3989 & 0.37   & p & G  & 14.2  & 4.36 &  0.8 & 0.87 & L  &  162.1 & 25.85 & P  &    \\
J1452$-$2638     & 223.0490 $-$26.6468 & 0.29   & p & G  & 12.5  & 3.26 &  2.7 & 1.23 & A  &   86.8 & 25.99 & MA &    \\
J1452$-$0230     & 223.1967 $-$02.5116 & 1.3450 & s & Q  &  6.54 & 3.30 &  5.0 & 0.64 & A  &    4.2 & 26.22 & A  &    \\
J1511$+$4605     & 227.8889 $+$46.0975 & 0.3467 & s & G  & 10.2  & 3.00 &  3.3 & 1.16 & L  &  204.5 & 25.89 & P  &    \\
J1526$+$2620$^c$ & 231.6449 $+$26.3342 & 0.1507 & s & G  & 28.0  & 4.40 &      &      & L  &        &       & P  &    \\
J1529$+$6015     & 232.3840 $+$60.2596 & 0.896  & s & G  & 13.2  & 6.16 &  0.8 & 0.92 & L  &   68.4 & 26.37 & P  &    \\
J1530$+$5243     & 232.6855 $+$52.7315 & 0.4460 & s & G  &  8.9  & 3.06 &  4.0 & 0.60 & L  &   57.0 & 25.58 & P  &    \\
J1542$+$3736     & 235.5374 $+$37.6055 & 0.6893 & s & Q  &  8.85 & 3.77 &  2.0 & 0.76 & L  &  211.9 & 26.59 & N  & i  \\
J1546$+$6838     & 236.5477 $+$68.6460 & 0.77   & p & G  &  8.28 & 3.68 &  0.0 & 0.95 & L  &   74.6 & 26.25 & P  & i  \\
J1547$+$5147     & 236.7916 $+$51.7857 & 1.1    & p & G  &  7.37 & 3.61 &  1.9 & 1.22 & L  &   94.1 & 26.71 & P  &    \\
J1558$-$2138     & 239.7356 $-$21.6401 & 0.55   & e & G? &  9.4  & 3.70 & 45.0 & 1.16 & R  & 3370.7 & 28.19 & n  &    \\
J1605$+$4434$^g$ & 241.3273 $+$44.5795 & 1.0    & p & G  &  7.2  & 3.46 &  1.4 & 0.51 & L  &  134.5 & 26.77 & N  &    \\
J1616$+$4825     & 244.0053 $+$48.4265 & 0.263  & s*& G  & 12.9  & 3.14 &  0.0 & 0.88 & L  &  805.9 & 26.21 & P  & i  \\
J1616$+$1111$^c$ & 244.0939 $+$11.1933 & 0.3574 & s & G  & 12.6  & 3.79 &  1.0 & 1.20 & R  &   47.0 & 25.91 & P  &    \\
J1623$-$6919     & 245.9165 $-$69.3210 & 0.53   & p & Qc &  8.1  & 3.06 &  3.8 & 0.43 & A  &  125.0 & 26.74 & A  &    \\
\end{tabular}}
\end{table*}

\begin{table*}

\begin{threeparttable}

\centering
\footnotesize{
\begin{tabular}{lclccrcccrcccl}
J1626$+$5434     & 246.7358 $+$54.5726 & 0.4886 & s & G  &  9.82 & 3.55 &  3.8 & 1.14 & L  &   73.2 & 25.78 & P  &    \\
J1636$+$5417     & 249.2461 $+$54.2904 & 0.5027 & s & G  &  8.45 & 3.10 &  3.0 & 1.19 & L  &  300.0 & 26.42 & P  & bi \\
J1637$+$3949$^h$ & 249.4737 $+$39.8211 & 0.75   & p & G  &  8.4  & 3.70 & 10.3 & 1.84 & L  &   40.4 & 25.96 & P  & i  \\
J1637$+$3204     & 249.4926 $+$32.0822 & 0.9    & p & G  &  7.82 & 3.66 &  4.2 & 1.46 & L  &   25.9 & 25.95 & P  &    \\
J1637$+$1936     & 249.4928 $+$19.6107 & 0.605  & s*& G  &  9.03 & 3.64 &  3.8 & 0.63 & R  &  130.6 & 26.88 & N  & m  \\
J1650$+$3912     & 252.5892 $+$39.2117 & 0.8774 & s & G  &  6.76 & 3.14 &  6.7 & 0.98 & L  &  102.4 & 26.52 & P  & i  \\
J1651$+$3209     & 252.8075 $+$32.1621 & 0.7454 & s & Q  &  7.4  & 3.29 &  0.7 & 0.66 & L  &  286.5 & 26.80 & P  &    \\
J1727$+$4615     & 261.9125 $+$46.2599 & 1.4    & p & G  &  6.55 & 3.31 &  2.0 & 0.71 & L  &  114.4 & 27.04 & N  &    \\
J1753$-$0332     & 268.4162 $-$03.5458 & 0.6    & p & GP & 13.6  & 5.46 &  1.0 & 0.67 & R  &   50.0 & 26.45 & A  &    \\
J1757$+$4051     & 269.3995 $+$40.8651 & 0.7    & p & G  & 10.45 & 4.48 &  2.0 & 0.75 & L  &   48.7 & 25.97 & P  & i  \\
J1801$+$5107$^c$ & 270.3239 $+$51.1229 & 0.50   & p & Qc & 10.0  & 3.66 &  5.0 & 1.23 & N  &   16.0 & 25.95 & P  &    \\
J1812$+$4409     & 273.0041 $+$44.1522 & 0.9    & p & Qc &  6.9  & 3.23 &  0.9 & 0.94 & L  &   66.2 & 26.36 & P  &    \\
J1812$+$6731     & 273.2044 $+$67.5170 & 0.95   & p & Qc &  6.55 & 3.11 & 12.2 & 0.46 & L  &   17.8 & 25.84 & ED &    \\
J1821$+$2639$^c$ & 275.3623 $+$26.6598 & 0.085  & s & G  & 39.0  & 3.70 &      &      & L  &        &       & P  & b  \\
J1841$+$4005     & 280.4619 $+$40.0927 & 1.05   & p & G  &  8.48 & 4.12 & 11.5 & 1.14 & R  &   91.0 & 27.28 & N  &    \\
J1902$-$5605     & 285.6145 $-$56.0914 & 0.37   & p & G  & 10.7  & 3.29 &  1.4 & 0.70 & A  &   53.0 & 26.01 & A  &    \\
J2035$-$6434     & 308.8202 $-$64.5742 & 0.72   & p & Qc & 10.6  & 4.60 &  1.0 & 1.38 & A  &   20.3 & 26.27 & A  & i  \\
J2048$-$0520     & 312.0116 $-$05.3361 & 0.52   & p & G  &  8.83 & 3.30 & 10.1 & 0.96 & R  &  100.4 & 26.62 & N  &    \\
J2108$-$3932     & 317.1487 $-$39.5499 & 0.35   & p & G  & 11.2  & 3.32 &  5.0 & 0.60 & R  &  211.0 & 26.54 & MA &    \\
J2129$-$5340     & 322.4761 $-$53.6700 & 1.26   & p & G  &  6.1  & 3.06 &  1.8 & 1.44 & A  &   37.9 & 27.01 & A  &    \\
J2206$+$2751     & 331.5236 $+$27.8501 & 0.3595 & s & G  & 13.0  & 3.92 &  4.6 & 1.06 & L  &   85.7 & 25.54 & P  &    \\
J2216$-$5737     & 334.2396 $-$57.6216 & 0.83   & p & G  &  7.4  & 3.37 &  0.0 & 0.81 & A  &   70.2 & 26.95 & A  &    \\
J2223$+$1831     & 335.8557 $+$18.5238 & 0.7316 & s & G  &  7.92 & 3.45 &  0.0 & 1.00 & L  &  167.5 & 26.55 & P  & i  \\
J2232$+$2857     & 338.1007 $+$28.9648 & 0.65   & p & G  &  7.55 & 3.14 &  7.0 & 0.95 & L  &   40.2 & 25.81 & P  &    \\
J2236$+$2512     & 339.2074 $+$25.2118 & 0.65   & p & G  &  7.5  & 3.12 &  2.0 & 1.14 & L  &  179.3 & 26.46 & P  &    \\
J2238$+$3305     & 339.5527 $+$33.0942 & 0.4896 & s & G  & 13.55 & 4.91 &  1.0 & 0.84 & L  &  145.3 & 26.08 & n  & b  \\
J2301$+$2401     & 345.3558 $+$24.0301 & 0.4    & p & G  &  9.58 & 3.09 &  0.0 & 0.78 & L  &  169.0 & 25.94 & P  & i  \\
J2321$-$1942     & 350.2988 $-$19.7149 & 0.582  & p & G  &  9.9  & 3.91 &  4.2 & 0.89 & R  &   38.1 & 26.30 & MW &    \\
J2347$-$6644     & 356.7532 $-$66.7419 & 0.86   & p & G  &  7.2  & 3.32 &  2.0 & 0.56 & A  &   15.6 & 26.33 & A  & m  \\
\end{tabular}}
\vspace{3mm}
\noindent 
\footnotesize{$^a$ GRGs J0101+5052 and J0843$-$0208 may have alternative hosts at lower redshift which would imply LLS values of 1.3 and 2.7\,Mpc, respectively. However, we consider the hosts
quoted here as the more likely ones.}\\
\footnotesize{$^b$ The host of this GRG was wrongly identified by
\citet{2023A&A...672A.163O} as SDSS~J013406.32+301537.2, but high-resolution
images at 1.4 and 5\,GHz by \citet{2019ApJS..241...37W} show it to be a
restarted RG hosted by SDSS~J013410.63+301537.7, also detected as X-ray source
4XMM~J013410.7+301538.}\\
\footnotesize{$^c$ These 7 sources are from \citet{2023A&A...672A.163O} but are not in the public 
LoTSS\,DR2 footprint. For 3 of these we could not find suitable other radio surveys 
to determine ALR, BA and $P_{145}$. Moreover, for J0722+4242 we adopt a host, SDSS~J072238.26+424227.2, different from that in \cite{2023A&A...672A.163O} since it coincides with a 0.7-mJy VLASS radio core.}\\
\footnotesize{$^d$ Listed as J0905+5527 in \citet{2023A&A...672A.163O} who propose SDSS~J090528.26+552711.7 
as the host, which is undetected in both LoTSS and VLASS. We propose the host
DESI~J136.4012+55.4346 as it has a radio core in both LoTSS and VLASS and is 
located closer to the GRG's major axis. Its higher redshift implies a higher LLS.}\\
\footnotesize{$^e$ Listed with $z_{\rm spec}$=1.8857 in \cite{2025ApJ...978..101D}
but the spectrum is very noisy and the redshift unreliable (M. Debski, priv.comm.)}\\
\footnotesize{$^f$ \citet{2023A&A...672A.163O} propose SDSS~J132002.53+602231.0 as the host which
lies off due SW of the GRG's major radio axis, is faint in LoTSS and undetected
in VLASS. We propose SDSS~J132024.00+602235.2 as the host, as it lies closer to
the major source axis and has jet-like emission on both sides directed towards
the outer lobes.}\\
\footnotesize{$^g$ We agree with the host proposed by \citet{Hardcastle2023} but note
that SDSS~J160513.70+443347.8 may be another candidate. Its WISE color, redshift,  
and thus LLS, would be similar but its ALR$\sim$1.05 would make
it statistically less probable.}\\
\footnotesize{$^h$ \citet{Hardcastle2023} list a smaller LAS of 6.4$'$ and propose
SDSS~J163749.12+395038.3 with a strong compact radio source
as the host. The DESI Legacy Survey DR10 resolves the SDSS object 
into a close pair of galaxies $\sim1.5''$ apart, and VLASS shows 
the fainter SW component to be the host of the radio source. 
Since this source is located off the GRG's major radio axis we prefer
SDSS~J163753.66+394916.0 as the host which is closer to the symmetry
centre of the GRG and has more AGN-like WISE colors. The redshift and
LLS are similar for both hosts.}

\end{threeparttable}

\end{table*}

\end{appendix}
  	
\end{document}